\documentclass[longauth]{aa} 
\usepackage{natbib}
\usepackage{graphicx}
\usepackage{lscape}
\usepackage{longtable}
\usepackage{multirow}
\usepackage[multiple]{footmisc}
\usepackage{txfonts}
\usepackage{color}                                                             

\begin{document}

   \title{Incidence of debris discs around FGK stars \\
          in the solar neighbourhood
    \thanks{{\it Herschel} is an ESA space observatory with science instruments 
            provided by European-led Principal Investigator consortia and 
            with important participation from NASA.
            } 
    }


   \author{B. Montesinos\inst{1,3}
          \and C. Eiroa\inst{2,3} 
          \and A. V. Krivov\inst{4}
          \and J. P. Marshall\inst{5,6}
          \and G. L. Pilbratt\inst{7}
          \and R. Liseau\inst{8}
          \and A. Mora\inst{9}
          \and J. Maldonado\inst{10}
          \and S. Wolf\inst{11}
          \and S. Ertel\inst{12}
          \and A. Bayo\inst{13,14}
          \and J.-C. Augereau\inst{15,16}
          \and A. M. Heras\inst{7}
          \and M. Fridlund\inst{8,17}
          \and W. C. Danchi\inst{18}
          \and E. Solano\inst{1,19}
          \and F. Kirchschlager\inst{11}
          \and C. del Burgo\inst{20}
          \and D. Montes\inst{21}
} 

\institute{Departmento de Astrof\'{\i}sica, Centro de Astrobiolog\'{\i}a 
          (CAB, CSIC-INTA), ESAC Campus, Camino Bajo del Castillo s/n, 
          E-28692 Villanueva de la Ca{\~n}ada, Madrid, Spain
\and Universidad Aut\'onoma de Madrid, Dpto. F\'isica Te\'orica, 
     M\'odulo 15, Facultad de Ciencias, Campus de Cantoblanco, 
     E-28049 Madrid, Spain
\and Unidad Asociada CAB--UAM
\and Astrophysikalisches Institut und Universit{\"a}tssternwarte,
     Friedrich-Schiller-Universit{\"a}t, Schillerg{\"a}{\ss}chen 2-3, 07745
     Jena, Germany
\and School of Physics, UNSW Australia, Sydaney NSW 2052, Australia
\and Australian Centre for Astrobiology, UNSW Australia, Sydney NSW
     2052, Australia
\and ESA, Directorate of Science, Scientific Support Office,
     European Space Research and Technology Centre (ESTEC/SCI-S),
     Keplerlaan 1, NL-2201 AZ Noordwijk, The Netherlands
\and Department of Earth and Space Sciences, Chalmers University of
     Technology, Onsala Space Observatory, Se-439 92 Onsala, Sweden
\and ESA-ESAC Gaia SOC, P.O. Box 78, E-28691 Villanueva de la
     Ca{\~n}ada, Madrid, Spain
\and INAF, Osservatorio Astronomico di Palermo, Piazza Parlamento 1,
     90134 Palermo, Italy
\and Institute of Theoretical Physics and Astrophysics,
     Christian-Albrechts University Kiel, Leibnizstr. 15, 24118 Kiel,
     Germany
\and Steward Observatory, Department of Astronomy, University of
     Arizona, 933 North Cherry Avenue, Tucson, AZ 85721, USA
\and Instituto de F\'{\i}sica y Astronom\'{\i}a, Facultad de Ciencias,
     Universidad de Valpara\'{\i}so, Av. Gran Breta\~na 1111, 5030 Casilla,
     Valpara\'{\i}so, Chile 
\and ICM Nucleus on Protoplanetary Disks, Universidad de Valpara\'{\i}so, 
     Av. Gran Breta\~na 1111, Valpara\'{\i}so, Chile
\and Univ. Grenoble Alpes, IPAG, F-38000 Grenoble, France 
\and CNRS, IPAG, F-38000 Grenoble, France 
\and Leiden Observatory, University of Leiden, PO Box 9513,
     2300 RA, Leiden, The Netherlands
\and NASA Goddard Space Flight Center, Exoplanets and Stellar Astrophysics, 
     Code 667, Greenbelt, MD 20771, USA
\and Spanish Virtual Observatory, Centro de Astrobiolog\'{\i}a (CAB, 
     CSIC-INTA), ESAC Campus, Camino Bajo del Castillo s/n, 
     E-28692 Villanueva de la Ca\~nada, Madrid, Spain
\and Instituto Nacional de Astrof\'{\i}sica, \'Optica y Electr\'onica, 
     Luis Enrique Erro 1, Santa Mar\'{\i}a de Tonantzintla, Puebla, M\'exico
     Departamento de Astrof\'{\i}sica, Facultad de Ciencias F\'{\i}sicas, 
     Universidad Complutense de Madrid, E-28040 Madrid
}

   \offprints{B. Montesinos \\  \email{benjamin.montesinos@cab.inta-csic.es}}
   \date{Received , ; Accepted  }

 
\abstract
{Debris discs are a consequence of the planet formation process and
  constitute the fingerprints of planetesimal systems.  Their solar
  system's counterparts are the asteroid and Edgeworth-Kuiper belts.}
{The aim of this paper is to provide robust numbers for the incidence
  of debris discs around FGK stars in the solar neighbourhood.}
{The full sample of 177 FGK stars with $d\!\leq\!20$ pc proposed for the
  DUNES survey is presented. {\it Herschel}/PACS observations at 100
  and 160 $\mu$m complemented in some cases with data at 70 $\mu$m,
  and at 250, 350 and 500 $\mu$m SPIRE photometry, were obtained. The
  123 objects observed by the DUNES collaboration were presented in a
  previous paper. The remaining 54 stars, shared with the DEBRIS
  consortium and observed by them, and the combined full sample are
  studied in this paper. The incidence of debris discs per spectral
  type is analysed and put into context together with other parameters
  of the sample, like metallicity, rotation and activity, and age.}
{The subsample of 105 stars with $d\!\leq\!15$ pc containing 23 F, 33
  G and 49 K stars, is complete for F stars, almost complete for G
  stars and contains a substantial number of K stars to draw solid
  conclusions on objects of this spectral type. The incidence rates
  of debris discs per spectral type are 0.26$^{+0.21}_{-0.14}$ (6
  objects with excesses out of 23 F stars), 0.21$^{+0.17}_{-0.11}$ (7
  out of 33 G stars) and 0.20$^{+0.14}_{-0.09}$ (10 out of 49 K
  stars), the fraction for all three spectral types together being
  0.22$^{+0.08}_{-0.07}$ (23 out of 105 stars). The uncertainties
  correspond to a 95\% confidence level.  The medians of the upper
  limits of $L_{\rm dust}/L_*$ for each spectral type are
  $7.8\times10^{-7}$ (F), $1.4\times10^{-6}$ (G) and
  $2.2\times10^{-6}$ (K); the lowest values being around
  $4.0\times10^{-7}$. The incidence of debris discs is similar for
  active (young) and inactive (old) stars. The fractional luminosity
  tends to drop with increasing age, as expected from collisional
  erosion of the debris belts.}
{}


\authorrunning{B. Montesinos et al.}
\titlerunning{Debris discs around nearby FGK main-sequence stars}

   \maketitle

\section{Introduction}
\label{Section:introduction}

Star and planet formation are linked by the presence of a
circumstellar disc built via angular momentum conservation of the
original molecular cloud undergoing gravitational collapse (e.g.
\citealt{Armitage2015}). These primordial discs, formed by gas and
small dust particles, evolve with the star during the pre-main
sequence (PMS) phase and, as they experience gas dispersal and grain
growth, are transformed from gas-dominated protoplanetary discs to
tenuous, gas-poor, dusty discs, so-called ``debris discs''. These
discs consist of $\mu$m-sized particles with short lifetimes, thus,
they need to be constantly replenished.  Hence, they are
second-generation discs, considered to be produced by the constant
attrition --due to collisional cascades-- of a population of
planetesimals (e.g. \citealt{Wyatt2008}, \citealt{Krivov2010}).

Primordial discs around young stars, as the birth-sites of planets,
provide the initial conditions for planet formation, the raw material
and the system's architecture. Some of these young discs have built
the $\sim\!2000$ exo-solar planets --at the time of writing this
paper-- distributed in several hundreds of multiple planetary systems
that have been found in the last 20 years, mostly around main-sequence
(MS) stars\footnote{See http://exoplanetarchive.ipac.caltech.edu/ for
  updated numbers of exoplanets and multi-planet systems.}. The
observed planetary systems as a whole, planets and/or debris discs,
present a large variety of architectures (e.g. \citealt{Marshall2014},
  \citealt{MoroMartin2015}, \citealt{Wittenmyer2015}), that in
  conjunction with the host stars, determine the fate of the systems
  once the stars abandon the MS phase towards later phases of stellar
  evolution (e.g. \citealt{Mustill2014}).

A large effort has been devoted to the field of debris discs, with
fundamental contributions from space astronomy in the mid- and
far-infrared ($\lambda\!\gtrsim\!10$ $\mu$m) since the pioneering
discovery of an infrared excess in the spectral energy distribution
(SED) of Vega provided by the {\it InfraRed Astronomical Satellite}
(IRAS) \citep{Aumann1984}. An enormous amount of references could be
provided to cover the vast area of debris discs, their properties,
morphology and relationship with the presence of planets.  Concerning
the topics addressed in this paper, the {\it Infrared Space
  Observatory} (ISO, 1995--1998) enlarged the sample of stars
observed, studying for the first time the incidence rate of debris
discs around MS AFGK stars \citep{Habing2001} and the time dependency
of Vega-like excesses \citep{Decin2003}. The {\it Spitzer} observatory
(cryogenic mission 2003--2009) provided qualitatively and
quantitatively large leaps forward, showing that $\sim\!16$ \% of
solar-type FGK stars have dusty discs \citep{Trilling2008}. The
sensitivity of these three missions to debris disc detection was
limited by their aperture sizes.

The ESA {\it Herschel} space observatory (2009--2013)
\citep{Pilbratt2010}, with its 3.5-m aperture, and imaging photometers
PACS \citep{Poglitsch2010}, and SPIRE \citep{Griffin2010}, providing
an increased sensitivity to debris discs (see Fig. 1 in
\citealt{Eiroa2013}), a wider wavelength coverage, and
  ability to spatially resolve many of them, has been fundamental in
  extending the picture. An account of results of the pre-{\it
    Herschel} era can be found in the review by \cite{Wyatt2008},
  whereas a much more comprehensive view, including some of the {\it
    Herschel} results is given by \cite{Matthews2014}.

\cite{Eiroa2013} (E13 herafter), presented the results obtained from
the {\it Herschel} Open Time Key Programme (OTKP)
DUNES\footnote{http://www.mpia-hd.mpg.de/DUNES/}\footnote{http://sdc.cab.inta-csic.es/dunes/},
DUst around NEarby Stars. This programme aimed to detect
Edgeworth-Kuiper belt (EKB) analogues around nearby solar-type
stars. The incidence rate of debris discs in a $d\!\leq\!20$ pc
subsample was $20\!\pm\!2$ \%. However, the sample of FGK stars
observed and analysed in that paper was not complete (see Section
\ref{Section:completeness} for details), therefore it is mandatory to
analyse data on a complete --or near-complete-- sample, in order to
avoid biased conclusions.

This paper analyses the full sample of solar-type (FGK) stars of the
DUNES programme located at $d\!\leq\!20$ pc, as it was described in
the original DUNES proposal, i.e.  the subsample studied in E13 plus
the stars shared with, and observed within, the OTKP DEBRIS, Disc
Emission via a Bias-free Reconnaissance in IR and Sub-mm
\citep{Matthews2010}. The scientific background, context and rationale
of the present work are the same as those presented in E13; they were
described in detail in Sections 1 and 2 of that paper and therefore
the information will not be duplicated again here. Special attention
is paid to the DUNES subsample of stars with $d\!\leq\!15$ pc,
which is complete for F stars, almost complete for G stars, and has a
substantial number of K stars, making the conclusions, concerning the
incidence rate of debris discs in the corner of our Galaxy, fairly robust.

The paper is organized as follows. In Sect. \ref{Section:sample}, the
sample of stars is described; the reasons why the observations of the
full DUNES sample were split between the DUNES and DEBRIS teams are
explained; the completeness of the sample is discussed, and
comprehensive information on the optical and near-IR photometry of the
shared sample used to build the spectral energy distributions is also
given. In Sect. \ref{Section:observations}, the observations and data
reduction are described. The results are presented in Sect.
\ref{Section:results}. The analysis of the full DUNES sample is done
in Sect. \ref{Section:analysis} and a summary of the main conclusions
is presented in Sect. \ref{Section:summary}.

\section{The stellar sample} 
\label{Section:sample}

\subsection{Selection criteria}
\label{Section:selection}

The stellar sample analysed in this work is the merger of the sample
studied in E13, which will be called DUNES\_DU in this paper, and a
subsample of FGK stars observed by the DEBRIS team, which will be
identified here as DUNES\_DB. The full DUNES sample is composed of the
merger of DUNES\_DU and DUNES\_DB. We will concentrate our analysis on
the stars with $d\!\leq\!20$ pc and pay special attention, when
addressing the incidence rates of debris discs, to the subset within
15 pc. Below, we give details of the DUNES\_DU and DUNES\_DB
subsamples.

\begin{table}[!ht]
\begin{center}
\caption{Summary of spectral types in the DUNES\_DU and DUNES\_DB samples.}
\label{Table:sample_spectraltypes}
\begin{tabular}{lcccc}
\hline\hline\noalign{\smallskip}
Sample                                 & F       &     G   & K       & Total   \\
\noalign{\smallskip}
\hline\noalign{\smallskip}
DUNES\_DU                               &  27     &    52   &   54    &  133    \\    
\noalign{\smallskip}
$\leq\!15$ pc DUNES\_DU subsample       &   4     &    19   &   43    &   66    \\
$\leq\!20$ pc DUNES\_DU subsample       &  19     &    50   &   54    &  123    \\
\noalign{\smallskip}
\hline\noalign{\smallskip}
DUNES\_DB                               &  51     &    24   &    6    &   81    \\
\noalign{\smallskip}
$\leq\!15$ pc DUNES\_DB subsample       &  19     &    14   &    6    &   39    \\
$\leq\!20$ pc DUNES\_DB subsample       &  32     &    16   &    6    &   54    \\
\noalign{\smallskip}
\hline\noalign{\smallskip}
\multicolumn{5}{l}{Note on the nomenclature of the samples:}\\
\noalign{\smallskip}
\multicolumn{5}{l}{DUNES\_DU contains the stars of the DUNES sample observed by}\\
\multicolumn{5}{l}{the DUNES team, DUNES\_DB contains the stars of the DUNES }\\
\multicolumn{5}{l}{sample observed by the DEBRIS team.}\\
\end{tabular}
\end{center}
\end{table} 

DUNES and DEBRIS were two complementary {\it Herschel} programmes with
different observing strategies and different approaches to the
selection criteria for their samples. Given the overlapping scientific
interests of both projects, the {\it Herschel} Time Allocation
Committee suggested to split the samples in the most convenient and
efficient way, keeping the philosophy and observing strategy of the
corresponding science cases. Both teams were granted with the same
amount of observing time (140 hours).

As described in E13, the original DUNES stellar sample, out of which
the final sample was built, was chosen from the {\it Hipparcos}
catalogue (VizieR online catalogue I/239/hip\_main,
\citealt{Perryman1997}) following the only criterion of selecting MS
--luminosity class V-IV/V-- stars closer than 25 pc, without any bias
concerning any property of the objects. The restriction to build the
final sample was that the stellar photospheric emission could be
detected by PACS at 100 $\mu$m with a S/N$\geq$5, i.e., the expected
100 $\mu$m photospheric flux should be significantly higher than the
expected background as estimated by the Herschel HSPOT tool at that
wavelength. Two stars, namely $\tau$ Cet (HIP 8102, G8 V) and
  $\epsilon$ Eri (HIP 16537, K2 V), although fulfilling all the
  selection criteria described above, do not belong to the DUNES
  sample because they were included in the Guaranteed Time Key
  Programme ``Stellar Disk Evolution'' (PI: G. Olofsson).

Taking into account the amount of observing time allocated and the
complementarity with DEBRIS, the sample observed by the DUNES team was
restricted to main-sequence FGK solar-type stars located at distances
shorter than 20 pc. In addition, FGK stars between 20 and 25 pc
hosting exoplanets (3 stars, 1 F-type and 2 G-type, at the time of the
proposal writing) and previously known debris discs, mainly from the
{\it Spitzer} space telescope (6 stars, all F-type) were also
included. Thus, the sample of stars directly observed by DUNES, which
we call DUNES\_DU in this paper\footnote{Note that this subsample was
  formally called the ``DUNES sample'' in E13, but it is actually only
  the subset of the full DUNES sample that was observed within the
  DUNES observing time.}, is composed of 133 stars, 27 out of which
are F-type, 52 G-type, and 54 K-type stars. The 20 pc subsample was
formed by 123 stars --19 F-type, 50 G-type and 54 K-type\footnote{In
  E13, the sample of stars within 20 pc was composed of 124 objects
  because the selection of the initial sample was done according to
  the original ESA 1997 release of the {\it Hipparcos} catalogue; in
  that release the parallax of HIP 36439 was $50.25\!\pm\!0.81$
  mas. However, in the revision by \cite{VanLeeuwen2007} the parallax
  is $49.41\!\pm\!0.36$ mas, putting this object beyond 20 pc. Here
  we exclude this star from the DUNES\_DU 20-pc sample,
  and consider only 123 stars in that subset.}.

The OTKP DEBRIS was defined as a volume limited study of A
through M stars selected from the ``UNS'' survey \citep{Phillips2010},
observing each star to a uniform depth, i.e., DEBRIS was a
flux-limited survey.  In order to optimize the results according to
the DUNES and DEBRIS scientific goals, the complementarity of both
surveys was achieved by dividing the common stars of both original
samples, considering whether the stellar photosphere could be detected
with the DEBRIS uniform integration time.  Those stars were assigned
to be observed by DEBRIS. In that way, the DUNES observational
requirement of detecting the stellar photosphere was satisfied. The
few common A- and M-type stars in both surveys were also assigned to
DEBRIS.

The net result of this exercise was that 106 stars observed by DEBRIS
satisfy the DUNES photospheric detection condition and are, therefore,
shared targets. Specifically, this sample comprises 83 FGK stars --51
F-type, 24 G-type and 8 K-type (the remaining stars are of spectral
type A and M). Note that spectral types listed in the {\it Hipparcos}
catalogue are used to give these numbers, we will see below that two K
stars had to be excluded due to a wrong spectral type
classification. Since the assignment to one of the teams was made on
the basis of both DUNES and DEBRIS original samples, the number of
shared targets located closer than 20 pc, i.e., the revised DUNES
distance, is fewer: 56 FGK - 32 F-type, 16 G-type, and 8 K-type stars.

\begin{figure}[!htbp]
\includegraphics[scale=0.5]{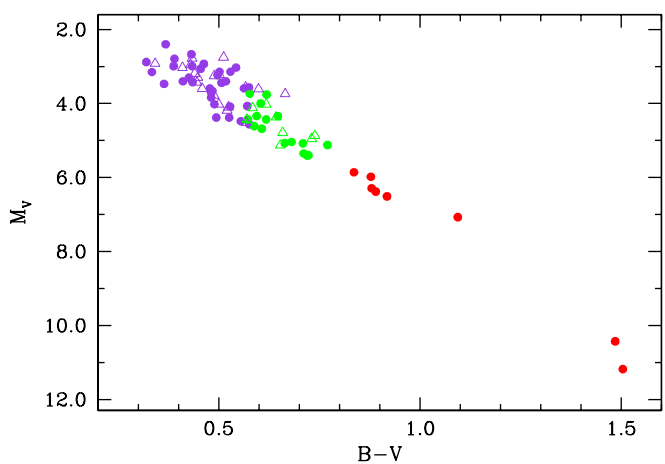}
\caption{HR diagram for the FGK stars of the DUNES sample observed
  by the DEBRIS team (DUNES\_DB). Colour codes -following the {\it
    Hipparcos} spectral types are: F stars (violet), G stars (green),
  K stars (red). Solid circles represent FGK stars closer than 20 pc;
  empty triangles are FGK stars further away. The two red dots 
  at the bottom right of the diagram are two stars classified as K5 
  in the {\it Hipparcos} catalogue but their colours correspond to M 
  stars. See text for details.}
\label{Figure:hrdiagram}
\end{figure}

In Fig. \ref{Figure:hrdiagram} we show the HR diagram
$M_V$--$(B\!-\!V)$ for the sample of FGK stars of the DUNES sample
observed during the DEBRIS observing time. Solid circles represent FGK stars
closer than 20 pc --i.e. the subsample analysed in this paper-- whereas
empty triangles have been used to plot FGK stars further away,
not included in the study presented here. Two stars, namely HIP 84140
and HIP 91768 --the two red dots at the bottom right of the diagram--
both are classified as K5 in the {\it Hipparcos} catalogue, but their
$(B\!-\!V)$ colours clearly correspond to that of an $\sim$M3 star (the
spectral type listed in SIMBAD for both objects), therefore they have
been moved out of the FGK star sample. We identify the subsample of 81
FGK stars as DUNES\_DB in this paper, out of which 54 are closer than 20
pc. Table \ref{Table:sample_spectraltypes} summarizes the spectral
type distribution of the DUNES\_DU and DUNES\_DB samples, taking into
account these two misclassifications. 

Table \ref{Table:stars_sample} provides some basic information of the
FGK stars within $d\!\leq\!20$ pc of the DUNES\_DB
sample\footnote{Excerpts of Tables 2, 3, 5 and 7 are included in the
  main body of the manuscript. The full tables can be found at the end
  of the paper and are available in electronic format.} (the
corresponding information for the DUNES\_DU sample can be found in
Table 2 of E13). Columns 1, 2 and 3 give the {\it Hipparcos}, and HD
numbers, and an alternative denomination. {\it Hipparcos} spectral
types are given in column 4. In order to check the consistency of
these spectral types we have explored VizieR using the DUNES discovery
tool\footnote{http://sdc.cab.inta-csic.es/dunes/searchform.jsp} (see
Appendix A in E13).  Results of this exploration are summarized in
column 5 which gives the spectral type range of each star taken into
account SIMBAD, \cite{Gray2003,Gray2006}, \cite{Wright2003} and the
compilation made by \cite{Skiff2009}. The typical spectral type range
is 2-3 subtypes.  Columns 6 and 7 list the equatorial coordinates and
columns 8 and 9 give parallaxes with errors and distances,
respectively.  The parallaxes are taken from \cite{VanLeeuwen2007}
(VizieR online catalogue I/311).  Parallax errors are typically less
than 1 mas. Only one star, namely HIP 46509, has an error larger than
2 mas, this object being a spectroscopic binary (see Table 4).  The
multiplicity status of the DUNES\_DB sample is addressed in
Sect. \ref{Section:multiplicity}.

\subsection{Completeness}
\label{Section:completeness}

The main constraint affecting the completeness of the sample is the
observational restriction that the photosphere had to be detected with a
S/N$\geq$5 at 100 $\mu$m. It is important to know the impact of
this restriction in order to assess the robustness of the results
concerning the frequency of the incidence of debris discs in the 
solar neighbourhood.

\begin{figure}[!htbp]
\includegraphics[scale=0.70]{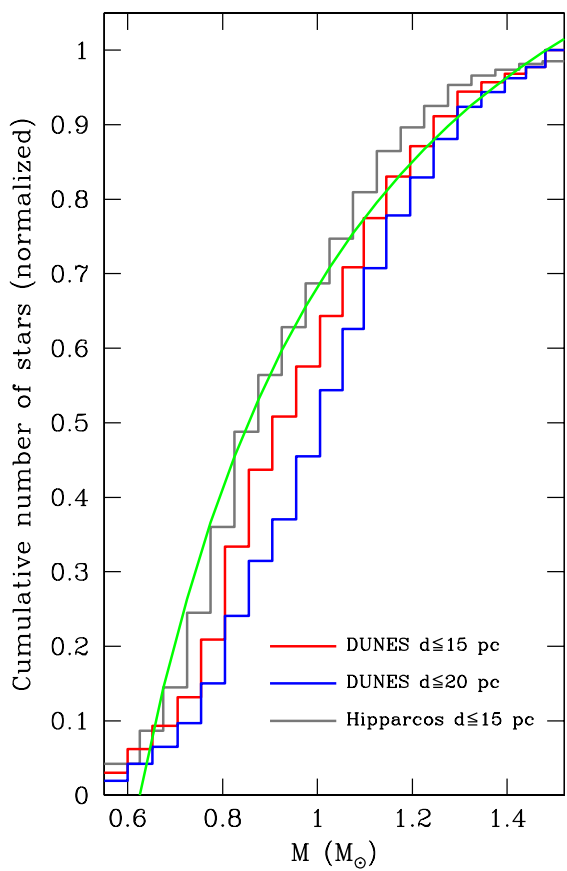}
\caption{Cumulative distributions (normalized) for the DUNES samples
  with $d\!\leq\!15$ pc (105 stars, red) and $d\!\leq\!20$ pc (177
  stars, blue) and the FGK stars in the {\it Hipparcos} catalogue with
  $d\!\leq\!15$ pc (154 stars, grey), all plotted against stellar
  mass. A Salpeter law with exponent $-2.35$ has been normalized to
  the distributions. See text for details.}
\label{Figure:completitude}
\end{figure}

In Fig. \ref{Figure:completitude} the cumulative numbers of stars in
the merged DUNES\_DU and DUNES\_DB $d\!\leq\!15$-pc and
$d\!\leq\!20$-pc samples --normalized to the total number of objects
in each one-- have been plotted against stellar mass as red and blue
histograms, respectively.  The masses have been assigned to each star
by linear interpolation of the values given for FGK MS stars by
\cite{Pecaut2013}\footnote{See also ``Stellar Color/Teff Table'' under
  section ``Stars'' in http://www.pas.rochester.edu/$\sim$emamajek/}
using $(B\!-\!V)$ as the independent variable. The expected mass
spectrum for the solar neighbourhood according to a Salpeter law
\citep{Salpeter1955} with exponent $-2.35$ has been normalized to the
total cumulative number of each sample and plotted in green.  It is
clear that the sample with $d\!\leq\!20$ pc shows an underabundace of
stars below $\sim$1.2 M$_\odot$ whereas the behaviour of the sample
with $d\!\leq\!15$ pc approximates that of the Salpeter law, but still
runs slighly below at low masses. Similar results are obtained using
other approaches for the mass spectrum, e.g. that by
\cite{Kroupa2001}. In grey, the cumulative distribution for the FGK
stars with $d\!\leq\!15$ pc from the {\it Hipparcos} catalogue has
been also included, showing a much closer agreement with the Salpeter
law.

Since the DUNES sample was drawn from the {\it Hipparcos} catalogue,
it is important to assess the problem of its completeness.  In
Sect. 2.2 of \cite{Turon1992} it can be seen that the {\it Hipparcos}
Input Catalogue was only complete down to $V\!=\!7.3$ for stars cooler
than G5. For $d\!=\!15$ pc, this corresponds to an spectral type of
$\sim$K2. In addition, not all the stars in the Input Catalogue were
observed (see Table 1 of \citealt{Turon1992}): for the magnitude bin
$V\!:\!7\!-\!8$ the efficiency of the survey was 93\%, therefore, even
applying a safety margin of 0.3 mag, we would have completeness only
up to spectral type K1.  In summary, down to 15 pc, {\it Hipparcos} is
complete for F and G but not for K stars.

\begin{figure}[!htbp]
\includegraphics[scale=0.90]{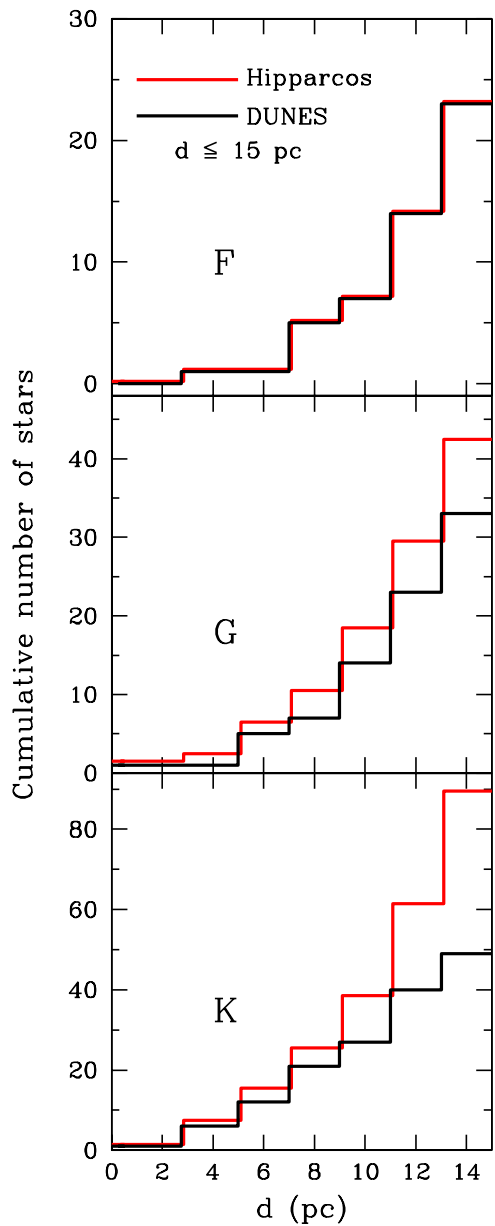}
\caption{Cumulative number of F, G and K stars (luminosity classes V
  and IV-I) in the {\it Hipparcos} catalogue (red) and in the DUNES
  sample (black), both at $d\!\leq\!15$ pc.}
\label{Figure:comparison}
\end{figure}

To quantify the departure of the full DUNES $d\!\leq\!15$-pc sample
from the parent {\it Hipparcos} set, we have compared the number of
stars of each spectral type in our sample with the corresponding
numbers in the {\it Hipparcos} catalogue. In
Fig. \ref{Figure:comparison}, cumulative histograms of the numbers of
F, G, and K stars (luminosity classes V and IV-V) in the {\it
  Hipparcos} catalogue (red), and in the DUNES sample (black), both at
$d\!\leq\!15$, are plotted against distance. The total numbers of
stars are 23 F, 42 G, 89 K in {\it Hipparcos} against 23 F, 33 G, 49 K
in the DUNES sample, i.e., the observational constraint imposed in the
final selection of the targets keeps all the F stars from {\it
  Hipparcos} but discards 9 G and 40 K stars. Therefore, a correction
on the fraction of excesses extracted from the DUNES sample would have
to be applied --in particular for K stars-- in order to give results
referred to a wider sample.  We will come back to this point in
Sect. \ref{Section:results}.


\begin{table*}[!htbp]
\begin{small}
\setcounter{table}{2}
\begin{tabular}{lllllllrr}
\multicolumn{9}{c}{{\bf \tablename\ \thetable{}.} The subsamples of DUNES\_DB with $d\!\leq\!15$ pc and $15\!\,\,{\rm pc}\!<\!d\!\leq\!20$ pc (only the first stars are shown).}\\
\\
\hline\hline
HIP & HD & Other ID & SpT (Hipparcos) & SpT range & \multicolumn{2}{c}{ICRS (2000)} & \multicolumn{1}{c}{$\pi$(mas)} & \multicolumn{1}{c}{$d$(pc)} \\
\\                                             
\multicolumn{9}{l}{FGK stars ($d\leq15$  pc)} \\\hline
1599  & 1581   & $\zeta$ Tuc  & F9V & F9.5V, F9V--G0V & 00 20 04.260 & $-$64 52 29.25 &  116.46$\pm$0.16 &  8.59$\pm$0.01 \\       
3765  & 4628   & LHS 121      & K2V & K2.5V, K2V      & 00 48 22.977 & $+$05 16 50.21 &  134.14$\pm$0.51 &  7.45$\pm$0.03 \\       
7751  & 10360J & p Eri        & K0V & K0V, K1--K3V    & 01 39 47.540 & $-$56 11 47.10 &  127.84$\pm$2.19 &  7.82$\pm$0.13 \\    
\multicolumn{9}{l}{\dots}\\
\multicolumn{9}{l}{FGK stars ($15\,\,{\rm pc}\!<d\leq20$  pc)}\\\hline   
5862  &  7570  &  $\nu$ Phe   & F8V     & F9V, F9V        & 01 15 11.121 & $-$45 31 54.00  &  66.16$\pm$0.24 & 15.11$\pm$0.05 \\   
17651 &  23754 &  27 Eri      & F3/F5V  & F5IV-V, F3III   & 03 46 50.888 & $-$23 14 59.00  &  56.73$\pm$0.19 & 17.63$\pm$0.06 \\   
36366 &  58946 &  $\rho$ Gem  & F0V...  & F0V, F0V--F3V   & 07 29 06.719 & $+$31 47 04.38  &  55.41$\pm$0.24 & 18.05$\pm$0.08 \\ 
\multicolumn{9}{l}{\dots}\\\hline
\end{tabular}
\end{small}
\label{Table:stars_sample}
\end{table*}


\subsection{Stellar parameters of the DUNES\_DB sample}

Table 3 gives some parameters of the DUNES\_DB objects with $d\!<\!20$
pc, namely the effective temperature, gravity, and metallicity
(cols. 2--4, and a flag in col. 5 indicating whether the determination
was spectroscopic or photometric), the stellar luminosity (col. 6) and
the activity indicator $\log R'_{\rm HK}$ (col. 7 and its
correspondent reference in col. 8). References for $T_{\rm eff}$,
$\log g$, [Fe/H] and $\log R'_{\rm HK}$ are given at the bottom of the
table. The luminosity was computed using equation (9) from
\cite{Torres2010}, with the values of the $V$-magnitude and colour
index $(B\!-\!V)$ taken from the {\it Hipparcos} catalogue, and the
bolometric corrections (BC) from \cite{Flower1996}; ($B\!-\!V$) was
used as the independent variable to obtain BC.

Col. 9 lists the rotation periods, $P_{\rm rot}$, estimated using the
strong correlation between $\log R'_{\rm HK}$ and the Rossby number,
defined as ${\rm Ro}\!=P_{\rm rot}/\tau_{\rm c}$, where $\tau_{\rm c}$
is the convective turnover time, which is a function of the spectral
type (colour) (\citealt{Noyes1984}, \citealt{Montesinos2001},
\citealt{Mamajek2008}). For six objects, the estimation of $P_{\rm rot}$
was not feasible due to the fact that the calibrations do not hold
for the values of either $\log R'_{\rm HK}$ or $B\!-\!V$; in that case,
lower limits of the rotation period, based on the values of $v \sin i$,
are given. Cols. 10 and 11 show the stellar ages $t_{\rm Gyro}$ and
$t_{\rm HK}$ computed according to gyrochronology and using the
chromospheric emission as a proxy for the age. In Appendix A we give
details of the estimation of the rotation periods and ages.

\subsection{Multiplicity}
\label{Section:multiplicity}

Table 4 lists the multiplicity properties of the objects that are
known to be binaries. The importance of this information cannot be
underestimated, specially when the spectral types of the components
are close --and hence $\Delta V$ is small-- and the binary is spectroscopic or
its components have not been resolved individually during a particular
spectroscopic or photometric observation from which a given parameter
is derived. In many cases parameters are assigned as representative of
the whole system, without reference in the literature to which
component they refer to. This has also an impact on the calculation of
the corresponding model photospheres that are used, after
normalization to the optical and near-IR photometry, as baselines to
detect potential excesses at the {\it Herschel}
wavelengths. Concerning the data given in Table 4, this has to be
taken into account for HIP 7751 ($\Delta V=0.16$), HIP 44248 (1.83),
HIP 61941 (0.04), HIP 64241 (0.68), HIP 72659 (2.19), HIP 73695
(0.87), and HIP 75312 (0.37), as well as for those stars classified as
``spectroscopic binaries'' for which no information on their
components is available. The proper motions of the components have
been extracted from the ``{\it Hipparcos} and {\it Tycho} catalogues (Double and
Multiples: Component solutions)'' (VizieR online catalogue
I/239/h\_dm\_com).

The bottom part of Table 4 gives information about the stars of the
DUNES\_DB sample with $d\!\leq\!20$ pc that, at the time of writing
this paper, have confirmed exoplanets. The mass and the
length of the semimajor axis of the orbit for each planet are given. 

\subsection{Photometry}

Tables 5a--5d give the optical, near-IR, AKARI, WISE, IRAS and {\it
  Spitzer} MIPS magnitudes and fluxes of the DUNES\_DB stars with
$d\!\leq\!20$ pc, and the corresponding references. This photometry
has been used to build and trace the spectral energy distributions
(SEDs) of the stars, that are, along with PHOENIX/Gaia models
\citep{Brott2005}, the tools to predict the fluxes at the far-IR
wavelengths in order to determine whether or not the observed fluxes
at those wavelengths are indicative of a significant excess. Appendix
C in E13 gives details on the photospheric models and the
normalization procedure to the observed photometry. In addition
  to the photometry, {\it Spitzer}/IRS spectra of the targets, built
  by members of the DUNES consortium from original data from the {\it
    Spitzer} archive, are included in the SEDs, although they have not
  been used in the process of normalization of the photospheric
  models.


\begin{table*}[!htbp]
\begin{center}
\begin{small}
\setcounter{table}{3}
\begin{tabular}{lcccccccccc}
\multicolumn{11}{c}{{\bf \tablename\ \thetable{}.} Parameters of the DUNES\_DB stars with $d\!\leq\!15$ pc and $15\!\,\,{\rm pc}\!<\!d\!\leq\!20$ pc (only the fist stars are shown).}\\
\\
\hline\hline
HIP & $T_{\rm eff}$ & $\log g$ & [Fe/H] & Flag &  $L_{\rm bol}$ & $\log R'_{\rm HK}$ & Ref & $P_{\rm rot}$ (d)  & $t_{\rm Gyro)}$ & $t_{\rm HK}$\\
    &  (K)          & (cm/s$^2$)    & (dex) &      & ($L_\odot$)   &                   &     & or lower limit\dag& (Gyr)            &  (Gyr)        \\
\hline
\hline
\multicolumn{11}{l}{ }\\
\multicolumn{11}{l}{FGK stars ($d\!\leq\!15$  pc)} \\\hline                 
1599    & 5960 & 4.45 & $-0.17$ & S  & 1.224  & $-4.855$ & 1 & $14.01\pm0.97$        & 2.20   & 4.03   \\ 
3765    & 4977 & 4.57 & $-0.22$ & S  & 0.283  & $-4.852$ & 2 & $39.02\pm0.25$        & 5.41   & 3.98   \\ 
7751    & 4993 & 4.54 & $-0.23$ & S  & 0.304  & $-4.94 $ & 3 &                       &        &        \\ 
7918    & 5891 & 4.36 & $+0.06$ & S  & 1.405  & $-4.987$ & 4 & $21.10\pm0.12$        & 3.56   & 6.36   \\  
7981    & 5189 & 4.52 & $-0.04$ & S  & 0.439  & $-4.912$ & 2 & $39.60\pm0.43$        & 6.04   & 4.99   \\ 
\hline
\end{tabular}
\end{small}
\end{center}
\end{table*}



\begin{table*}[!htbp]
\begin{center}
\begin{small}
\setcounter{table}{5}
\begin{tabular}{lrlllll}
\multicolumn{7}{c}{{\bf \tablename\ \thetable{a}.} Johnson $V$, $B\!-\!V$, Cousins $ V\!-\!I$ and Str\"omgren photometry.}\\
\\
\hline
\hline
HIP  &  \multicolumn{1}{c}{$V$} & \multicolumn{1}{c}{$B\!-\!V$} & \multicolumn{1}{c} {$V\!-\!I$} & \multicolumn{1}{c}{$b\!-\!y$} &  
\multicolumn{1}{c}{$m_1$} & \multicolumn{1}{c}{$c_1$} \\   
     &  \multicolumn{1}{c}{(mag)} & \multicolumn{1}{c}{(mag)} & \multicolumn{1}{c} {(mag)} & \multicolumn{1}{c}{(mag)} &  
\multicolumn{1}{c}{(mag)} & \multicolumn{1}{c}{(mag)} \\   

\hline
\multicolumn{7}{l}{ } \\
\multicolumn{7}{l}{FGK stars ($d\!\leq\!15$ pc)} \\\hline
  1599 & 4.23 & 0.576$\pm$0.010 & 0.65$\pm$0.02 & 0.368$\pm$0.003 & 0.177$\pm$0.004 & 0.302$\pm$0.020  \\
  3765 & 5.74 & 0.890$\pm$0.008 & 0.97$\pm$0.02 & 0.512$\pm$0.003 & 0.423$\pm$0.002 & 0.255$\pm$0.004  \\
  7751 & 5.76 & 0.880$\pm$0.400 & 0.93$\pm$0.02 & 0.512           & 0.421           & 0.262            \\
  7918 & 4.96 & 0.618$\pm$0.001 & 0.67$\pm$0.03 & 0.389$\pm$0.000 & 0.198$\pm$0.005 & 0.348$\pm$0.011  \\
  7981 & 5.24 & 0.836$\pm$0.008 & 0.88$\pm$0.01 & 0.492$\pm$0.002 & 0.367$\pm$0.007 & 0.296$\pm$0.006  \\
\hline
\label{ }
\end{tabular}
\setcounter{table}{5}
\begin{tabular}{lrrrcrrrrrrl}
\multicolumn{12}{c}{{\bf \tablename\ \thetable{b}.}  2MASS $JHK_{\rm s}$ and ancillary Johnson $JHKLL'M$ photometry.}\\
\\
\hline
\hline
HIP  &  \multicolumn{1}{c}{2MASS $J$} & \multicolumn{1}{c}{2MASS $H$} & \multicolumn{1}{c}{2MASS $K_{\rm s}$} 
&  Qflag & \multicolumn{1}{c}{$J$} & \multicolumn{1}{c}{$H$} & \multicolumn{1}{c}{$K$} &  
\multicolumn{1}{c}{$L$} & \multicolumn{1}{c}{$L'$} & \multicolumn{1}{c}{$M$} & Refs.      \\   
     &  \multicolumn{1}{c}{(mag)} & \multicolumn{1}{c}{(mag)} & \multicolumn{1}{c}{(mag)} 
&    &  \multicolumn{1}{c}{(mag)} & \multicolumn{1}{c}{(mag)} & \multicolumn{1}{c}{(mag)} &  
\multicolumn{1}{c}{(mag)} & \multicolumn{1}{c}{(mag)} & \multicolumn{1}{c}{(mag)} &     \\
\hline
\multicolumn{12}{l}{ }\\
\multicolumn{12}{l}{FGK stars ($d\!\leq\!15$)} \\\hline
  1599 &   3.068$\pm$0.272 &  2.738$\pm$0.218 &  2.769$\pm$0.250 & DDD      &  3.196 &  2.880 &  2.832 &  2.803 &  2.795 &  2.856 & 1,2,3 \\
  3765 &   4.367$\pm$0.310 &  3.722$\pm$0.230 &  3.683$\pm$0.268 & DDD      &        &        &        &        &        &        &     \\
  7751A&                   &                  &  3.558$\pm$0.270 &\multicolumn{1}{r}{D} &     &        &        &        &        &        &     \\
  7751B&   4.043$\pm$0.378 &                  &  3.510$\pm$0.282 &D$\;\;\,$D&        &        &        &        &        &        &     \\
  7918 &   4.000$\pm$0.262 &  3.703$\pm$0.226 &  3.577$\pm$0.314 & DDD      &  3.800 &  3.560 &        &        &        &        & 2    \\
  7981 &   3.855$\pm$0.240 &  3.391$\pm$0.226 &  3.285$\pm$0.266 & DDD      &  3.775 &  3.345 &  3.285 &        &        &        & 2    \\
\hline
\label{ }
\end{tabular}
\setcounter{table}{5}
\begin{tabular}{lcccccc}
\multicolumn{7}{c}{{\bf \tablename\ \thetable{c}.}  AKARI 9 and 18 $\mu$m fluxes and WISE W1, W3 and W4 photometry.}\\
\\
\hline
\hline
HIP   & \multicolumn{2}{c}{AKARI}   & &     \multicolumn{3}{c}{WISE}         \\\cline{2-3}\cline{5-7} 
      & 9  $\mu$m & 18  $\mu$m  & & 3.35 $\mu$m (W1) & 11.56 $\mu$m (W3) & 22.09
 $\mu$m (W4) \\
      & (mJy)      & (mJy)      & & (mag)            &  (mag)            &  (mag)            \\
\hline
\multicolumn{7}{l}{ } \\
\multicolumn{7}{l}{FGK stars ($d\!\leq\!15$ pc)} \\\hline 
1599   &   4191$\pm$28  &  1055$\pm$30  &   &                   &  2.856$\pm$0.011 &   2.788$\pm$0.019    \\                          
3765   &   2199$\pm$15  &   519$\pm$8   &   &                   &  3.370$\pm$0.039 &   3.493$\pm$0.023    \\                          
7751   &   3431$\pm$198 &   892$\pm$20  &   &                   &                  &   3.095$\pm$0.053    \\                          
7751B  &   1810$\pm$135 &               &   &                   &                  &                      \\
7918   &   2280$\pm$24  &   494$\pm$12  &   &  3.489$\pm$0.423  &  3.469$\pm$0.012 &   3.444$\pm$0.018    \\                          
7981   &   2724$\pm$5.  &   586$\pm$39  &   &  3.274$\pm$0.479  &                  &   3.312$\pm$0.033    \\ 
\hline
\label{ }
\end{tabular}
\setcounter{table}{5}
\begin{tabular}{ll@{ }rl@{ }rl@{ }rc@{}cc}
\multicolumn{10}{c}{{\bf \tablename\ \thetable{d}.} IRAS 12, 25, 60 $\mu$m and {\it Spitzer}/MIPS 24
 and 70 $\mu$m fluxes.}\\
\\
\hline
\hline
HIP   &\multicolumn{6}{c}{IRAS} & & \multicolumn{2}{c}{MIPS}  \\\cline{2-7}\cline{9-10}
      & 12 $\mu$m& \%   & 25 $\mu$m & \% & 60 $\mu$m & \%    & & \multicolumn{1}
{c}{24 $\mu$m}  & 
\multicolumn{1}{c}{70 $\mu$m} \\
      & \multicolumn{1}{c}{(mJy)} & &  \multicolumn{1}{c}{(mJy)} & & \multicolumn{1}{c}{(mJy)}& & & 
\multicolumn{1}{c}{(mJy)}  & \multicolumn{1}{c}{(mJy)}  \\
\hline
\multicolumn{10}{l}{ }\\
\multicolumn{10}{l}{FGK stars ($d\!\leq\!15$ pc)} \\
\hline
1599   & 3.11$\times10^3$ &  4  & 6.95$\times10^2$ &   4  &     &     & & 519$\pm$11 & 82$\pm$8 \\
3765   & 1.56$\times10^3$ &  7  &                  &      &     &     & & 264$\pm$5  & 26$\pm$10\\
7751   & 2.70$\times10^3$ &  5  & 6.55$\times10^2$ &   7  &     &     & & 231$\pm$4  & 22$\pm$4 \\ 
7918   & 1.67$\times10^3$ &  5  & 3.87$\times10^2$ &   8  &     &     & & 289$\pm$6  & 25$\pm$11\\
7981   & 1.87$\times10^3$ &  7  & 6.38$\times10^2$ &  10  &     &     & & 334$\pm$7  & 48$\pm$9 \\
\hline
\label{ }
\end{tabular}
\end{small}
\end{center}
\end{table*}



\section{\textit{\textbf{Herschel}} observations of the DUNES\_DB sources 
and data reduction}
\label{Section:observations}

\subsection{PACS observations}

As described by \cite{Matthews2010}, DEBRIS is a flux-limited survey
where each target was observed to a uniform depth (1.2 mJy beam$^{-1}$
at 100 $\mu$m). The observations of the DUNES\_DB sources were
performed by the DEBRIS team with the ESA {\it Herschel} Space
Observatory using the instrument PACS. Further details can be found in
the paper by \cite{Matthews2010}. The main parameters of the
observational set up used by DEBRIS are compatible with those
used by DUNES and described in E13.

\subsection{Data reduction}

Data for each source comprise a pair of mini scan-map observations
taken with the PACS 100/160 waveband combination.  DEBRIS and DUNES
observations both followed the same pattern for mini scan-map
observations (see PACS Observer’s Manual, Chapter 5), with minor
differences in the scan leg lengths (8\arcmin~for DEBRIS,
10\arcmin~for DUNES) leading to slightly different coverage across the
resulting map areas for the two surveys, in addition to the different
map depths. These differences are immaterial to the analysis presented
here.  The \textit{Herschel} PACS observations were reduced in the
Herschel Interactive Processing Environment \citep{Ott2010}. For the
analysis presented here we used HIPE version 10 and PACS calibration
version 45.  Reduction was carried out following the same scheme as
presented in E13. Table 6 shows a log of the observations with details
of the OBSIDs and the on-source integration times.

The individual PACS scans of each target were high-pass filtered to
remove large-scale background structure, using high pass filter radii
of 20 frames at 100~$\mu$m and 25 frames at 160~$\mu$m, suppressing
structures larger than 82\arcsec~and 102\arcsec~in the final images,
respectively. For the filtering process, a region 30\arcsec~radius
around the source position in the map along with regions where the
pixel brightness exceeded a threshold defined as twice the standard
deviation of the non-zero flux elements in the map (determined from
the level 2 pipeline reduced product) were masked from inclusion in
the high pass filter calculation. Regions external to the central
30\arcsec~that exceeded the threshold were tagged as source to be
avoided in the background estimation phase of the flux extraction
process. Deglitching was carried out using the spatial deglitching
task in median mode and a threshold of 10-$\sigma$. For the image
reconstruction, a drop size ({\tt pixfrac} parameter) of 1.0 was used.
Once reduced, the two individual PACS scans in each waveband were
mosaicked to reduce sky noise and suppress 1/$f$ striping effects from
the scanning. Final image scales were 1\arcsec~per
pixel at 100~$\mu$m and 2\arcsec~per pixel at
160~$\mu$m, compared to native instrument pixel sizes of 3.2\arcsec~for
100~$\mu$m, and 6.4\arcsec~for 160~$\mu$m.


\begin{table*}
\begin{small}
\setcounter{table}{7}
\begin{center}
\begin{tabular}{llrrrrrrl}
\multicolumn{9}{c}{{\bf \tablename\ \thetable{}.} PACS flux densities for the DUNES\_DB stars with $d\!\leq\!15$ pc and $15\!\,\,{\rm pc}\!<\!d\!\leq\!20$ pc (only the first stars are shown).}\\
\\
\hline
\hline
HIP & HD & \multicolumn{1}{c}{PACS100} &  \multicolumn{1}{c}{S100}  & $\chi_{100}$ & \multicolumn{1}{c}{PACS160} & \multicolumn{1}{c}{S160} & $\chi_{160
}$ & Status \\ 
    &    &  \multicolumn{1}{c}{(mJy)}  &  \multicolumn{1}{c}{(mJy)} &              & \multicolumn{1}{c}{(mJy)}   & \multicolumn{1}{c}{(mJy)} &          
   &        \\
\hline
\multicolumn{9}{l}{ }\\                                                         
\multicolumn{9}{l}{FGK stars ($d\!\leq\!15$  pc)} \\\hline
1599                  &  1581    &  33.23$\pm$ 2.35 & 30.08$\pm$ 0.19  & $ 1.34$ &   29.03$\pm$3.46  &  11.75$\pm$ 0.07 & $ 4.99$ & Dubious\\
3765                  &  4628    &  19.43$\pm$ 1.93 & 15.84$\pm$ 0.36  & $ 1.83$ &          $<3.38$  &   6.19$\pm$ 0.14 & $     $ & No excess \\\noalign{\vspace{0.05cm}}
\multirow{2}{*}{7751} &  10360   &  11.95$\pm$ 1.70 & 12.39$\pm$ 0.21  & $-0.26$ &          $<5.00$  &   4.84$\pm$ 0.08 & $     $ & No excess \\ 
                      &  10361   &  10.66$\pm$ 1.68 & 14.90$\pm$ 0.26  & $-2.49$ &          $<5.00$  &   5.82$\pm$ 0.10 & $     $ & No excess \\\noalign{\vspace{0.05cm}}
7918                  &  10307   &  14.33$\pm$ 1.75 & 16.21$\pm$ 0.25  & $-1.06$ &   10.35$\pm$3.17  &   6.33$\pm$ 0.10 & $ 1.27$ & No excess \\
7981                  &  10476   &  20.18$\pm$ 1.91 & 20.05$\pm$ 0.21  & $ 0.07$ &    7.87$\pm$2.05  &   7.83$\pm$ 0.08 & $ 0.02$ & No excess \\
\hline
\end{tabular}
\end{center}
\end{small}
\end{table*}


\section{Results}
\label{Section:results}

The extraction of PACS photometry of the sources analysed in this
paper follows identical procedures as those in E13. Sect. 6.1.1 of
that paper contains a very detailed description of the procedures to
carry out the flux extraction. Summarizing, flux densities were
measured using aperture photometry. If the image was consistent with a
point source, then an aperture appropriate to maximise the
signal-to-noise was adopted (i.e. 5\arcsec~ at 100~$\mu$m, and
8\arcsec~at 160~$\mu$m, see E13).  Regarding extended sources, we
refer the reader to Sect. 7.2.3 of E13, where the criteria to assess
whether the 3$\sigma$ flux contours in the 100 and 160 $\mu$m images
denote extended emission are described.

The background and rms scatter were estimated from the mean and
standard deviation of 10 square sky apertures placed at random around
the image. The sky apertures were sized to have roughly the same area
as the source aperture. The locations of the sky apertures were chosen
such that they did not overlap with the source aperture or any
background sources (to avoid contamination), and lay within
60\arcsec~of the source position (to avoid higher noise regions at the
the edges of the image).

Measured flux densities in each PACS waveband were corrected for the
finite aperture size using the tabulated encircled energy fractions
given in \cite{Balog2014}, but have not been colour
corrected\footnote{See PACS Technical Notes PICC-ME-TN-038 and
  PICC-ME-TN-044.}.  A point-source calibration uncertainty of 5\% was
assumed for all three PACS bands. Note that aperture corrections valid
for point sources have been also applied to the case of extended
sources. This is obviously an approach, but it should be valid as long
as the aperture size is larger than the extended source. This
criterion has widely been used, without any apparent inconsistency, in
other works, e.g. \cite{Wyatt2012} for 61 Vir, \cite{Duchene2014} for
$\eta$ Crv, \cite{Roberge2013} for 49 Cet.

Table 7 lists the PACS 100 and 160 $\mu$m photometry and 1$\sigma$
uncertainties for the sources of the DUNES\_DB sample with
$d\!\leq\!20$ pc, identified as the far-IR counterparts of the optical
stars. Uncertainties include both the statistical and systematic
errors added in quadrature; the uncertainty in the calibration is 5\% in 
both bands \citep{Balog2014}. Quantities without errors in the PACS160
column correspond to 3$\sigma$ upper limits. In the columns adjacent
to the PACS photometry, the photospheric predictions,
$S_\nu(\lambda)$, with the corresponding uncertainties, are given. The
fluxes $S_\nu(\lambda)$ are extracted from a Rayleigh-Jeans
extrapolation of the fluxes at 40 $\mu$m from the PHOENIX/Gaia
normalized models.  The uncertainties in the individual photospheric
fluxes were estimated by computing the total $\sigma$ of the
normalization, in logarithmic units; in that calculation, the observed
flux at each wavelength involved in the normalization process was
compared with its corresponding predicted flux. The normalized model
$\log S_\nu(\lambda)$ was permitted to move up and down a quantity
$\pm\sigma$. That value of $\sigma$ was then translated into
individual --linear-- uncertainties of the fluxes,
$\sigma(S_\nu(\lambda))$, at the relevant {\it Herschel} wavelengths.
The typical uncertainty in the photospheric predictions derived from a
change of $\pm 50$ K in $T_{\rm eff}$ for a model with 6000 K amounts
$\sim$0.9\% of the predicted flux density. It has not noticeable
effects in any of the calculations.

We consider that an object has an infrared excess at any of the 
PACS wavelengths when the significance:

\begin{equation}
\chi_\lambda=\frac{{\rm PACS}_\nu(\lambda) - S_\nu(\lambda)}
           {(\sigma({\rm PACS}_\nu(\lambda))^2+\sigma(S_\nu(\lambda))^2)^{1/2}} > 3.0
\label{Equation:chi}
\end{equation}

\noindent although in some cases the complexity of the fields makes it
difficult to ascertain whether an apparent clear excess based solely
on the value of $\chi_\lambda$ is real or not.

The values of $\chi_{\rm 100}$ and $\chi_{\rm 160}$ are listed in
Table 7 along with the status ``Excess/No excess'' or ``Dubious'' in
the last column, derived both from those numbers and a careful
inspection of the fields.

In addition, PACS 70 $\mu$m and SPIRE photometry from the literature
were included for HIP 61174, HIP 64924 and HIP 88745, and PACS 70
$\mu$m photometry for HIP 15510. The fluxes and the corresponding
references are given at the bottom of Table 7.

\begin{figure}[!ht]
\includegraphics[scale=0.85]{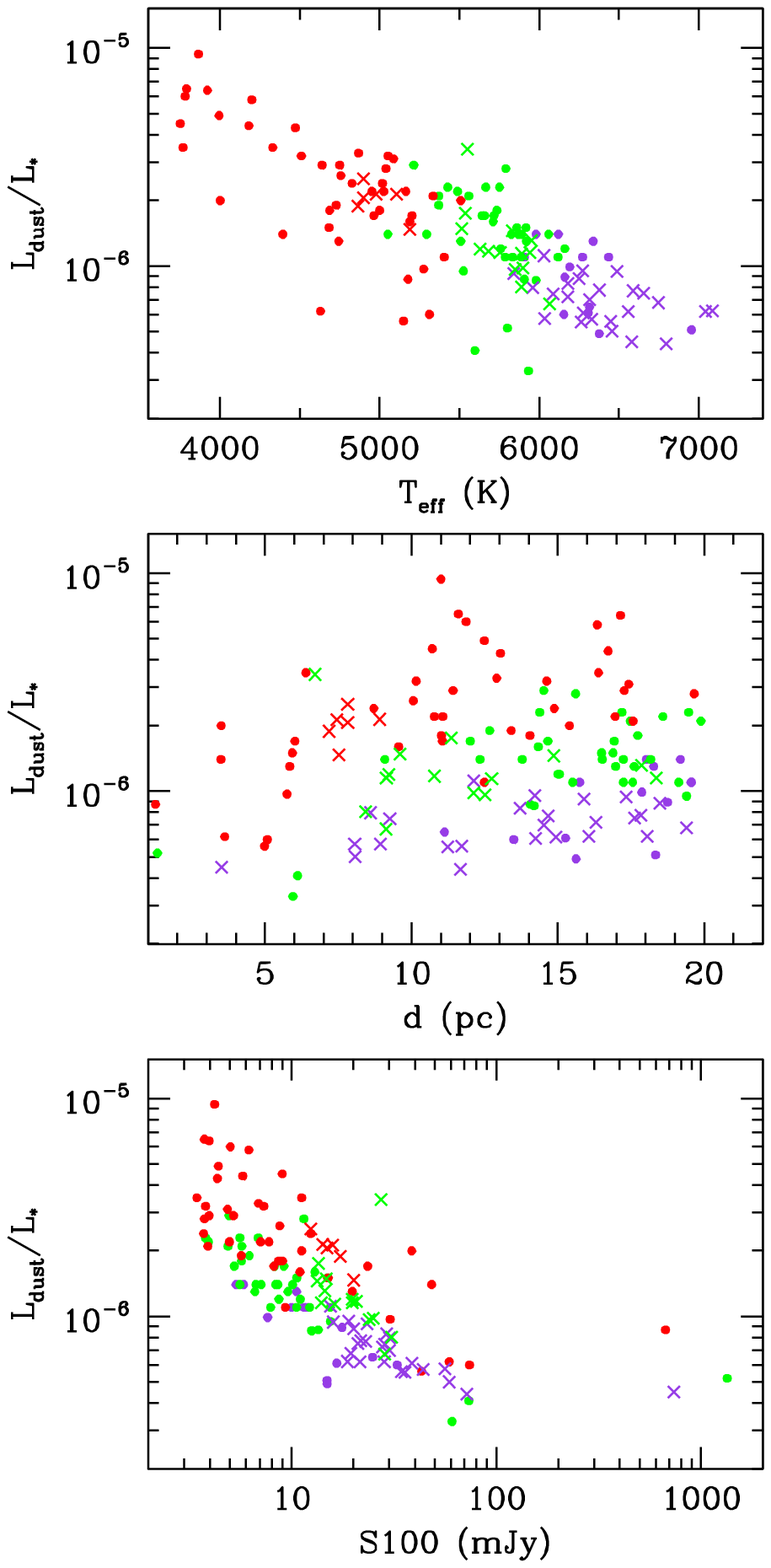}
\caption{Upper limits of the fractional luminosity of the dust for the
  non-excess sources of the DUNES\_DB (crosses) and the DUNES\_DU
  (dots) samples within 20 pc, plotted against the effective
  temperature, distance and stellar flux at 100 $\mu$m. Violet, green
  and red colours represent F, G and K stars. See text for details.}
\label{Figure:upper_limits}
\end{figure}

\subsection{Stars with excesses}
\label{Section:excesses}
  
After a careful analysis of the corresponding images and surrounding
fields, 11 stars of the DUNES\_DB sample with $d\!\leq\!20$ pc are
considered as showing real excesses among those in Table 7 having
$\chi_{100}$ and/or $\chi_{160}\!>\!3.0$. They are listed in
the upper half of Table \ref{Table:ldust_lstar}.

Eight sources show excess at both 100 and 160 $\mu$m; these are HIP
15510, HIP 16852, HIP 57757, HIP 61174, HIP 64924, HIP 71284, HIP
88745 and HIP 116771.

Two sources, HIP 5862 and HIP 23693, show excess only at 100
$\mu$m and fluxes at 160 $\mu$m slightly above the photospheric flux
but not large enough to give $\chi_{160}\!>\!3.0$. Consequently they
present a decrease in their SED for $\lambda\!\geq\!100$
$\mu$m, similar to those in three stars in DUNES\_DU (HIP 103389,
HIP 107350 and HIP 114948). An extensive modelling of these  three
objects was done by \cite{Ertel2012} and their conclusions can be
applied to HIP 5862 and HIP 23693.

HIP 113283 shows excess only at 160 $\mu$m, therefore it is a
cold-disc candidate as those discovered and modelled by
\cite{Eiroa2011}, E13, \cite{Krivov2013} and \cite{Marshall2013}.

Among the 11 sources with confirmed excesses, five have been spatially
resolved. They are marked with an asterisk in Tables 7 and
\ref{Table:ldust_lstar}. Four of them were already known, namely HIP
15510 (\citealt{Wyatt2012}, \citealt{Kennedy2015}), HIP 61174
(\citealt{Matthews2010},\citealt{Duchene2014}), HIP 64924
\citep{Wyatt2012} and HIP 88745 \citep{Kennedy2012}.

The fifth extended source found in this work is HIP 16852.  A
2-dimensional Gaussian fit gives full-widths at half-maximum (FWHM)
along the principal axes of 7.7\arcsec$\times$8.5\arcsec at 100
$\mu$m, and 12.2\arcsec$\times$10.5\arcsec at 160 $\mu$m, where those
of the PSF along the same directions at each wavelength are
6.9\arcsec$\times$6.8\arcsec and 11.8\arcsec$\times$10.9\arcsec,
respectively. This implies that the source is marginally resolved at
100 $\mu$m. Thus, HIP 16852 is similar in terms of angular size to HIP
15510, which was also marginally resolved at 100 $\mu$m by
\cite{Kennedy2015}\footnote{An inspection of the image at 70 $\mu$m
from the 70/160 waveband combination, that has not be used in this
work, shows that HIP 16852 is also extended at this wavelength.}.

Among the excess sources, planets have been detected around HIP 15510
\citep{Pepe2011}, and HIP 64924 \citep{Vogt2010}, see details in Table
4.

In addition to the excess sources, nine more objects have $\chi_{100}$
and/or $\chi_{160}\!>\!3.0$ but are labelled in Table 7 as ``Dubious''
because of the complexity of the fields that makes an unambiguous
measurement of the fluxes difficult. They are listed in the lower
  half of Table \ref{Table:ldust_lstar}. These are HIP 1599, HIP 36366,
HIP 44248, HIP 37279, HIP 59199, HIP 61941 (all of them F-type), HIP
72659, HIP 73695 (binaries with a primary of G-type) and HIP 77257
(G). 

In Table \ref{Table:ldust_lstar}, we give the results for $T_{\rm
  BB}$, estimated from black body fittings, and the fractional
luminosity $L_{\rm dust}/L_*$ computed by integrating directly on one
hand the black body curve that best fits the excess, and on the other,
the normalized photospheric model. Values are given both for the stars
with confirmed and apparent excesses. Typical uncertainties in the
black-body temperatures are $\pm\!10$ K. For all the stars, apart from
HIP 61174, the data involved in the black-body fittings are only the
PACS --and SPIRE, if available-- flux densities. For HIP 61174, the
fit to the warm black body includes all the ancillary photometry
between 10 and 30 $\mu$m (WISE W3 and W4, Akari 18, MIPS 24, IRAS 12
and 25) collected by us and listed in Tables 5c and 5d.

The values of the fractional luminosity that appear in the table for
HIP 5862 and HIP 23693 correspond to modified black body fits to all
the points showing excess, however, those fits have been used just to
integrate the excess fluxes, since the values of $\beta$ --the
parameter modifying the pure black body in
$(\lambda_0/\lambda)^\beta$-- are unreasonably high in both cases, of
the order of $\sim\!6.0$, making a physical interpretation very
difficult; most probably a two-ring disc would be more adequate to
model those two cases. The black body radii computed according to the
expression

\begin{equation}
R_{\rm BB}\,=\left(\frac{278}{T_{\rm BB}{\rm (K)}}\right)^2\,
            \left(\frac{L_*}{L_\odot}\right)^{1/2}
\label{Equation:rbb}
\end{equation}

\noindent (see e.g. \citealt{Backman1993}) are also included 
in the table.

Figs. B.1 and B.2 show the SEDs of the sources with confirmed and
dubious excesses, respectively. In blue we have plotted all the
photometry available below 70 $\mu$m, in red the PACS --and SPIRE for
some objects-- fluxes, in black the model photosphere normalized, and
in green the black body fits to the excess.

Information of some of these sources has been already reported 
in other works.  \cite{Gaspar2013} compiled a catalogue of {\it
  Spitzer}/MIPS 24 and 70 $\mu$m, and {\it Herschel}/PACS 100 $\mu$m
observations that includes the stars analysed in this paper. They
confirm excesses for seven out of the 11 stars listed at the top of
Table \ref{Table:ldust_lstar}, the exceptions being HIP 57757
($\chi_{100}\!=1.01$) and HIP 71284 ($\chi_{100}\!=\!3.35$, but no
MIPS data available); HIP 113283 is obviously labelled as
``no-excess'' by these authors because they did not analyse the
observation at 160 $\mu$m, the wavelength where we detect an
excess. For HIP 116771 they do not provide the PACS 100 $\mu$m flux
and the star is catalogued as a non-excess source according to the
MIPS 70 $\mu$m ($\chi_{70}\!=\!1.77$).

\setcounter{table}{7}
\begin{table}[!ht]
\caption{Black body dust temperatures and fractional dust luminosities for
  the DUNES\_DB stars with excesses.}
\setlength{\tabcolsep}{4.5pt}
\label{Table:ldust_lstar}
\begin{tabular}{lllccr}
\hline\hline\noalign{\smallskip}
HIP & HD   & SpT       & $T_{\rm BB}$ & $R_{\rm BB}$ &  \multicolumn{1}{c}{$L_{\rm dust}/L_*$}  \\
    &      &           &    (K)      &    (au)     &                     \\
\hline
    &      &           &           &              &       \\
\multicolumn{6}{l}{Stars with confirmed excesses}\\\hline\noalign{\smallskip}
5862    & 7570   & F8 V      &   96     &  12    & $2.4\times10^{-6}$ \\  
15510 * & 20794  & G8 V      &   32     &  60    & $6.3\times10^{-7}$ \\
16852 * & 22484  & F9 V      &   87     &  18    & $8.4\times10^{-6}$ \\
23693   & 33262  & F7 V      &   82     &  14    & $2.7\times10^{-6}$ \\ 
57757   & 102870 & F8 V      &   38     & 100    & $8.8\times10^{-7}$ \\
61174 * & 109085 & F2 V      &   40     & 107    & $2.2\times10^{-5}$ \\ 
64924 * & 115617 & G5 V      &   50     &  28    & $2.4\times10^{-5}$ \\ 
71284   & 128167 & F3 V      &  111     &  11    & $1.1\times10^{-5}$ \\
88745 * & 165908 & F7 V      &   45     &  52    & $1.1\times10^{-5}$ \\
113283  & 216803 & K4 Vp     &   --   &  --   & ${\it <2.0\times10^{-6}}$ \\
116771  & 222368 & F7 V      &   51     &  54    & $1.1\times10^{-6}$ \\
        &        &           &          &         &                    \\
\multicolumn{6}{l}{Stars with $\chi_{100}$ and/or  $\chi_{160}\!>\!3.0$
considered dubious excesses}\\\hline\noalign{\smallskip}         
1599   & 1581   & F9 V      &   18     & 264  & $9.6\times10^{-7}$ \\  
36366  & 58946  & F0 V...   &   21     & 405  & $5.7\times10^{-7}$ \\
37279  & 61421  & F5 IV-V   &  113     &  16  & $1.5\times10^{-6}$ \\
44248  & 76943  & F5 V      &   27     & 243  & $4.6\times10^{-7}$ \\
59199  & 105452 & F0 IV-V   &   21     & 358  & $6.9\times10^{-7}$ \\
61941  & 110379J& F0 V+...  &   30     & 248  & $3.3\times10^{-7}$ \\
72659  & 131156 & G8 V+K4 V & $\sim$40 & 40   & $2.9\times10^{-6}$ \\
73695  & 133640 & G2 V+G2 V &   23     & 182  & $1.7\times10^{-6}$ \\
77257  & 141004 & G0 V      &   33     & 102  & $7.6\times10^{-7}$ \\\noalign{\smallskip}
\hline\noalign{\smallskip}
\end{tabular}
\tablefoot{* denotes that the source is extended. Modified
  black bodies were fitted for HIP 5862 and HIP 23693 to compute these
  excesses (see text for details); the $T_{\rm BB}$ listed in the
  table correspond to pure black bodies fitted excluding the PACS
  fluxes at 160 $\mu$m, the corresponding $L_{\rm dust}/L_*$ for those
  temperatures being $5.22\times10^{-6}$ and $4.42\times10^{-6}$
  respectively. For HIP 61174 two black bodies were fitted to the warm
  and cold excesses, the warm excess corresponds to a $T_{\rm
    BB}\!=\!300$ K, which implies $L_{\rm
    dust}/L_*\!=\!1.65\times10^{-4}$. For HIP 113283 --a cold
    disc candidate-- there is only one excess flux density at 160
    $\mu$m, therefore a black body fit is not feasible; the value in
    italics for the upper limit of $L_{\rm dust}/L_*$ is just
    orientative and would correspond to a black body of 20 K
    normalized to the flux at 160 $\mu$m.}
\end{table} 

\cite{Chen2014} analysed and modelled the SEDs of 499 targets that
show {\it Spitzer}/IRS excesses, including a subset of 420 targets
with MIPS 70 $\mu$m observations. They found that the SEDs for the
majority of objects ($\sim\!66$\%) were better described using a
two-temperature model with warm ($T_{\rm dust}\!\sim\!100-500$ K) and
cold ($T_{\rm dust}\!\sim\!50-150$ K) dust populations analogous to
zodiacal and Kuiper Belt dust. Four of their excess sources for which
we also find an excess in the far-IR appear in \cite{Chen2014} (VizieR
online catalogue J/ApJS/211/25): HIP 5862, HIP 23693 and HIP 64924 are
modelled with two-temperature fits, the cold components having $T_{\rm
  dust}\!=\!94\!\pm\!8$, $51\!\pm\!7$ and $54\!\pm\!10$ K,
respectively, whereas HIP 16852 is reproduced by a one-temperature fit
with $T_{\rm dust}\!=\!100\!\pm\!6$ K.  The single black-body fits for
those stars (see Table \ref{Table:ldust_lstar}) are 96, 82, 50 and 87
K, respectively, which do not deviate much from the determinations by
Chen et al.

\cite{MoroMartin2015} also reported the detection/non-detection of
dust around stars of the DUNES and DEBRIS samples, based solely on 
their measurements of the {\it Herschel}/PACS 100 $\mu$m fluxes. Eight
out of the 11 stars we claim as having excess match their positive
detections, the exception being HIP 57757 for which they find a 
$\chi_{100}\!=\!2.43$, and HIP 113283, for which we only detect 
excess at 160 $\mu$m, as pointed out above; HIP 88745 does not appear 
in their sample.

Since the observed excess emission of some disc candidates is
relatively weak, one cannot exclude the possibility of contamination
by background galaxies (E13,
\citealt{Krivov2013}). \cite{Sibthorpe2013} carried out a cosmic
variance independent measurement of the extragalactic number counts
using PACS 100 $\mu$m data from the DEBRIS survey.  To estimate the
probability of galaxy source confusion in the study of debris discs,
Table 2 of that paper gives the probabilities of one background source
existing within a beam half-width half-maximum radius (3.4\arcsec for
PACS 100, 5.8\arcsec for PACS 160) of the measured source location for
a representative range of excess flux densities.
 
We have computed the excess flux densities at 100 and 160 $\mu$m for
the 11 discs in Table \ref{Table:ldust_lstar}. The smallest values at
each wavelength occur for HIP 15510 (8.0 mJy) and HIP 113283 (7.8
mJy). These values imply the probability of coincidental alignment
with a background object of 0.2\% and 1\%, respectively. Since the
excess flux densities are larger for the remaining targets, the
probabilities for them are obviously lower.

In the case of a survey, it is more relevant to ask what is the
probability that one or more members of the sample coincide with a
galaxy. We have computed the median 1$\sigma$ uncertainties --only
considering the rms of the measurement, not the calibration error-- of
the PACS 100 and 160 $\mu$m fluxes for the set of 54 stars of the
DUNES\_DB sample within 20~pc, the results being 1.63 and 3.18 mJy
respectively. Assuming fluxes for potential background sources of
three times those $\sigma$ values at the corresponding wavelengths, we
obtain at 100 $\mu$m a chance of coincidental alignment for a single
source in a sample of 54 stars of 1.1\%. This implies the following
probabilities $P(i)$ of having $i$ fakes in the sample: $P(0)$=55\%,
$P(1)$=33\%, $P(2)$=10\%, $P(3)$=2\% and $P(4)$=0.3\%. For 160 $\mu$m,
the chance of coincidental alignment of a single source is 2.9\%,
which implies $P(0)$=21\%, $P(1)$=33\%, $P(2)$=26\%, $P(3)$=13\% and
$P(4)$=5\%. Eq. (2) of the paper by \cite{Sibthorpe2013} and the
matrices provided in that work were used for computing these
estimates.

As we pointed out, five out of the 11 excess sources found in this
work are spatially resolved, therefore we can consider them as real
debris disc detections. According to the results above, the chances
that the remaining six sources were all contaminated by background
galaxies are $\sim0.003$\% and $\sim0.4$\% at 100 and 160 $\mu$m,
respectively.

\subsection{Stars without excesses}
\label{Section:noexcess}

Fig. B.3 shows the SEDs of the DUNES\_DB stars within 20 pc without
excesses. In Fig.  \ref{Figure:upper_limits} the $3\sigma$ upper
limits of the fractional luminosity of the dust for these sources are
plotted as crosses against the effective temperature, distance and
stellar flux at 100 $\mu$m. For the sake of completeness, the
non-excess DUNES\_DU sources within 20 pc have been also added (dots,
see Fig. 8 and Table 12 of E13). Violet, green and red colours
represent F, G and K stars. The dubious sources are not included. The
upper limits have been computed using the PACS 100 $\mu$m flux
  densities in the expression (4) by \cite{Beichman2006}, with a
representative $T_{\rm dust}\!=\!50$ K.

The plots in Fig. \ref{Figure:upper_limits} confirm and extend what
was already seen in E13. The upper panel shows that the upper limits
tend to decrease with $T_{\rm eff}$, the middle panel shows that for a
given distance the hotter the star the lower the upper limit, and the
lower panel shows that for a given predicted photospheric flux, which
depends both on the spectral type and the distance, the upper limits
increase with decreasing effective temperature. The general trend of
the $L_{\rm dust}/L_*$ upper limits decreasing with the stellar
temperature is expected just from pure black-body scaling
considerations and the construction of the DUNES survey. Finer
details, however, depend on the particular depth achieved for
different spectral types and each individual observation.

The lowest values of the upper limits for the fractional luminosity
reached are around $\sim\!4.0\times10^{-7}$, the median for each
spectral type been $7.8\times10^{-7}$, $1.4\times10^{-6}$ and
$2.2\times10^{-6}$ for F, G and K stars respectively; the median 
for the whole sample is $1.4\times10^{-6}$.

\begin{figure}[!ht]
\includegraphics[scale=0.70]{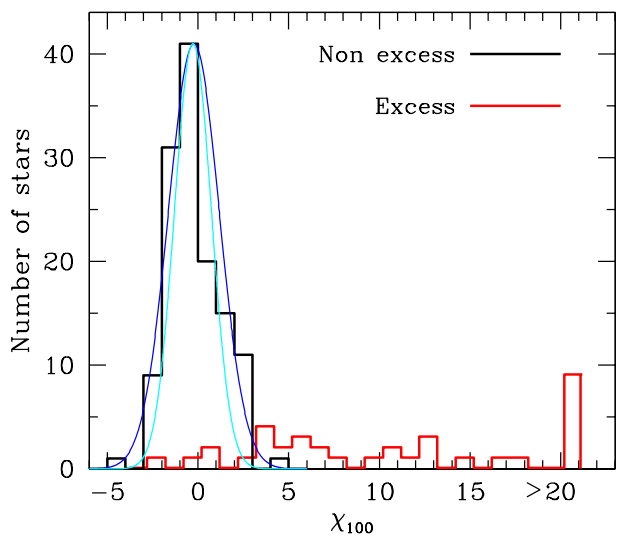}
\caption{Plotted in black, the histogram of the 100 $\mu$m
  significance $\chi_{100}$ for the non-excess sources of the merged
  DUNES\_DU and DUNES\_DB samples with $d\!\leq\!20$ pc (dubious
  sources of both samples are not included). A Gaussian with
  $\sigma\!=\!1.40$, the standard deviation of the $\chi_{100}$-values
  for those sources, is plotted in blue. In cyan, we show a Gaussian with
  $\sigma\!=\!1.02$, corresponding to the subsample of non-excess
  sources with $-2.0\!<\!\chi_{100}\!<2.0$. In red, the distribution
  of $\chi_{100}$ for the excess sources is also displayed. This
  figure is an extension of Fig. 6 of E13. See text for details.}
\label{Figure:chi100}
\end{figure}

In Fig. \ref{Figure:chi100} histograms of the significance
$\chi_{100}$ for the non-excess (black) and excess (red) sources of
the merged DUNES\_DU and DUNES\_DB samples with $d\!\leq\!20$ pc are
plotted. For the non-excess sources, the mean, median and standard
deviation of the $\chi_{100}$-values are $-0.23$, $-0.33$ and 1.40,
respectively. A Gaussian curve with $\sigma\!=\!1.40$ is also plotted.
These results are an extension of those already presented in Sect. 7.1
and Fig. 6 of E13, and are quantitatively very similar. The shift to
the peak of the distribution to a negative value of $\chi_{100}$
likely reflects the fact that the extrapolation of the model
photospheres to the PACS bands using a Rayleigh-Jeans approximation
does not take into account the decrease in brightness temperature at
$\sim\!100$ $\mu$m that occurs in late-type stars, and therefore
overestimates the photospheric emission around that wavelength. That
was directly observed using DUNES data for $\alpha$ Cen A
\citep{Liseau2013} and $\alpha$ Cen B \citep{Wiegert2014}. Note that
the stars in the red histogram with $\chi_{100}\!<\!3.0$ are those
considered as harbouring a cold disc, the excess being present only at
160 $\mu$m.

The fact that the standard deviation of the distribution departs from
1.0 --the expected value assuming a normal distribution-- poses a
problem already encountered in, e.g., E13 and \cite{Marshall2014}. One
potential origin of this could be the underestimation of the
uncertainties either in the PACS 100 $\mu$m flux or in the
photospheric predictions --or both-- that would have the effect of
increasing the absolute values of $\chi_{100}$, since the term
containing the uncertainties is in the denominator of the definition
of $\chi_\lambda$ (eqn. \ref{Equation:chi}). The distribution of
$\chi_{100}$ would become broader, and $\sigma$ would increase. We
have done the experiment of computing the mean, median and $\sigma$
for the non-excess sources considering only those stars with values of
$-2.0\!<\!\chi_{100}\!<\!2.0$, the goal being to test the central part
of the distribution. The results for the three quantities for this
subset of stars (108 objects) are $-0.28$ (mean), $-0.34$ (median) and
$\sigma\!=\!1.02$ (Gaussian curve plotted in cyan in
Fig. \ref{Figure:chi100}). This shows that the problem is more complex
than dividing the stars into two bins of ``excess'' and
``non-excess''. There must be a spectrum of values with an increasing
contribution from dust to the total measurement in the non-excess
sample until it hits a threshold and becomes an excess value, that has
been set in practice to $\chi_\lambda\!=\!3.0$. This can have the
effect of a departure of the full distribution from normal, while
narrowing the interval of $\chi_{100}$, ensuring that we deal with a
more conservative definition of non-excess sources, seems to have the
effect of approaching a normal distribution.
 
\begingroup
\begin{table*}[!htb]
\caption{Summary of stars per spectral type in the DUNES\_DU and DUNES\_DB samples and frequency of excesses.}
\label{Table:sample_excesses}
\setlength{\tabcolsep}{3.0pt}
\begin{tabular}{lccccccccccccccc}
\hline\hline\noalign{\smallskip}
  & \multicolumn{3}{c}{F}&& \multicolumn{3}{c}{G}&& \multicolumn{3}{c}{K} && \multicolumn{3}{c}{FGK}\\
\cline{2-4}\cline{6-8}\cline{10-12}\cline{14-16}\noalign{\smallskip}
\multirow{2}{*}{Sample}  & \multirow{2}{*}{T} & \multirow{2}{*}{E} & Frequency 
                        && \multirow{2}{*}{T} & \multirow{2}{*}{E} & Frequency 
                        && \multirow{2}{*}{T} & \multirow{2}{*}{E} & Frequency 
                        && \multirow{2}{*}{T} & \multirow{2}{*}{E} & Frequency\\
       &   &   & 95\% interval &&   &  & 95\% interval &&   &   & 95\% interval &&   &   & 95\% interval \\
\hline\noalign{\smallskip}
\multirow{2}{*}{$\leq\!15$ pc DUNES\_DU}             &  \multirow{2}{*}{4} & \multirow{2}{*}{2}  & 0.50 
                                                    && \multirow{2}{*}{19} & \multirow{2}{*}{5}  & 0.26 
                                                    && \multirow{2}{*}{43} & \multirow{2}{*}{9}  & 0.21 
                                                    && \multirow{2}{*}{66} & \multirow{2}{*}{16} & 0.24 \\
     &    &    & [0.15-0.85]    &&    &    &  [0.11-0.49]   &&    &    & [0.11-0.35] &&    &    &  [0.15-0.36] \\\noalign{\vspace{0.1cm}}
\multirow{2}{*}{$\leq\!20$ pc DUNES\_DU}            & \multirow{2}{*}{19} & \multirow{2}{*}{4}   & 0.21 
                                                   && \multirow{2}{*}{50} & \multirow{2}{*}{11}  & 0.22 
                                                   && \multirow{2}{*}{54} & \multirow{2}{*}{10}  & 0.19 
                                                   && \multirow{2}{*}{123}& \multirow{2}{*}{25}  & 0.20 \\
     &   &  & [0.08-0.44] &&    &    &  [0.13-0.35]   &&    &    & [0.10-0.31] &&     &    & [0.14-0.28]   \\\noalign{\vspace{0.15cm}}
\multirow{2}{*}{$\leq\!15$ pc DUNES\_DB}            & \multirow{2}{*}{19} & \multirow{2}{*}{4}  & 0.21 
                                                   && \multirow{2}{*}{14} & \multirow{2}{*}{2}  & 0.14 
                                                   && \multirow{2}{*}{6}  & \multirow{2}{*}{1}  & 0.17 
                                                   && \multirow{2}{*}{39} & \multirow{2}{*}{7}  & 0.18 \\
     &   &  & [0.08-0.44] &&   &  &  [0.03-0.41]  &&    &    & [0.01-0.58]  &&     &    & [0.09-0.33]  \\\noalign{\vspace{0.1cm}}
\multirow{2}{*}{$\leq\!20$ pc DUNES\_DB}            & \multirow{2}{*}{32} & \multirow{2}{*}{8}  & 0.25 
                                                   && \multirow{2}{*}{16} & \multirow{2}{*}{2}  & 0.13 
                                                   && \multirow{2}{*}{6}  & \multirow{2}{*}{1}  & 0.17 
                                                   && \multirow{2}{*}{54} & \multirow{2}{*}{11} & 0.20 \\
     &   &  & [0.13-0.42] &&   &  & [0.02-0.37]    &&   &    & [0.01-0.58]  &&     &    & [0.12-0.33]  \\\noalign{\vspace{0.15cm}}
\hline\noalign{\smallskip}
\multirow{2}{*}{$\leq\!15$ pc DUNES\_DU+DUNES\_DB}& \multirow{2}{*}{23}  & \multirow{2}{*}{6}  & 0.26        
                                                 && \multirow{2}{*}{33}  & \multirow{2}{*}{7}  & 0.21        
                                                 && \multirow{2}{*}{49}  & \multirow{2}{*}{10} & 0.20        
                                                 && \multirow{2}{*}{105} & \multirow{2}{*}{23} & 0.22 \\
     &   &   & [0.12-0.47] &&  &  & [0.10-0.38] &&  &  & [0.11-0.34] &&  &  & [0.15-0.31]\\\noalign{\vspace{0.1cm}}
\multirow{2}{*}{$\leq\!20$ pc DUNES\_DU+DUNES\_DB}& \multirow{2}{*}{51}  & \multirow{2}{*}{12} & 0.24        
                                                 && \multirow{2}{*}{66}  & \multirow{2}{*}{13} & 0.20        
                                                 && \multirow{2}{*}{60}  & \multirow{2}{*}{11} & 0.18        
                                                 && \multirow{2}{*}{177} & \multirow{2}{*}{36} & 0.20 \\
     &   &   & [0.14-0.37] &&  &  & [0.12-0.31] &&  &  & [0.10-0.30] &&  &  & [0.15-0.27]\\\noalign{\smallskip}
\hline\noalign{\smallskip}
\multirow{2}{*}{$\leq\!15$ pc {\it Hipparcos}}    & \multirow{2}{*}{23}  & \multirow{2}{*}{6} & 0.26        
                                                 && \multirow{2}{*}{42}  & \multirow{2}{*}{\it 9} & 0.21        
                                                 && \multirow{2}{*}{89}  & \multirow{2}{*}{\it 18} & 0.20        
                                                 && \multirow{2}{*}{154} & \multirow{2}{*}{\it 33} & {\it 0.21} \\
     &   &   & [0.12-0.47] &&  &  & {\it [0.11-0.36]} &&  &  & {\it [0.13-0.30]} &&  &  & {\it [0.16-0.29]}\\\noalign{\smallskip}
\hline\noalign{\smallskip}
\multicolumn{16}{l}{Note: ``T'' and ``E'' mean ``Total'' and ``Excess'', respectively.}
\end{tabular}
\label{Table:incidencerates}
\end{table*} 
\endgroup

\section{Analysis of the full DUNES sample}
\label{Section:analysis}

In E13 some conclusions were drawn based solely on the DUNES\_DU subsample. In 
this section we will comment on the relevant results that are 
derived from the analysis of the merged DUNES\_DU + DUNES\_DB sample, 
both with $d\!\leq\!15$ pc and $d\!\leq\!20$ pc.

\subsection{Excess incidence rates}
\label{Section:incidencerates}

Table 9 shows a summary of the excess incident rates in the DUNES\_DU,
DUNES\_DB, and the full sample. In particular, the $d\!\leq\!15$-pc
subsample contains 23 F, 33 G and 49 K stars. The incidence rates of
excesses are 0.26 (6 objects with excesses out of 23 F stars), 0.21 (7
out of 33 G stars) and 0. 20 (10 out of 49 K stars), the fraction for
the total sample with $d\!\leq\!15$ pc being 0.22 (23 out of 105
stars). Those percentages do not change significantly if we consider
all targets within $d\!\leq\!20$ pc. We also give the 95\% confidence
intervals for a binomial proportion for the corresponding counts
according to the prescription by \cite{Agresti1998}.

As specified in Sect. \ref{Section:noexcess}, the median of the upper
limits of the fractional luminosities per spectral type are
$7.8\times10^{-7}$, $1.4\times10^{-6}$ and $2.2\times10^{-6}$ for F, G
and K stars, respectively. As mentioned in E13, these numbers improve
the sensitivity reached by {\it Spitzer} by one order of
magnitude. Our results are to be compared with those found by
\cite{Trilling2008} (see Table 4 of their paper), where the incidence
rates of stars with excess at 70 $\mu$m, as measured by {\it
  Spitzer}/MIPS, are 0.18, 0.15 and 0.14 for F, G and K stars
respectively. The variation of the incidence rates with spectral type
is likely directly related to the dependence of the fractional
luminosity with $T_{\rm eff}$ (see Fig. \ref{Figure:upper_limits}).

As we showed in Sect. \ref{Section:completeness}, the merged sample
with $d\!\leq\!15$ pc analysed in this paper is complete for F stars,
almost complete for G stars, whereas a number of K stars are lost from
the parent set of the {\it Hipparcos} catalogue from which the final
DUNES sample was drawn, the reason being the observational constraint
imposed (background contamination, see Sect. \ref{Section:selection}).
In the lowest row of Table 9, we give the expected total number of
stars of each spectral type in the {\it Hipparcos} catalogue with
$d\!\leq\!15$ pc that would show excess, under the assumption that the
individual excess frequencies are those obtained from the full DUNES
sample; the numbers that are extrapolations for the G and K stars are
written in italics. The total incidence rate of excesses for the {\it
  Hipparcos} subset, namely $0.21^{+0.08}_{0.05}$, does not change
with respect to the DUNES result, $0.22^{+0.09}_{-0.07}$, the
confidence interval being obviously slightly narrower. Since both
$\tau$ Cet and $\epsilon$ Eri are within 15 pc and have debris discs
(\citealt{Lawler2014}, \citealt{Greaves2014}), should these sources be
included in the statistics, the total incidence rate of excesses
within $d\!\leq\!15$ pc would be $0.23^{+0.09}_{-0.07}$ (25 out of 107
objects).

\begin{figure}[!htbp]
\includegraphics[scale=0.65]{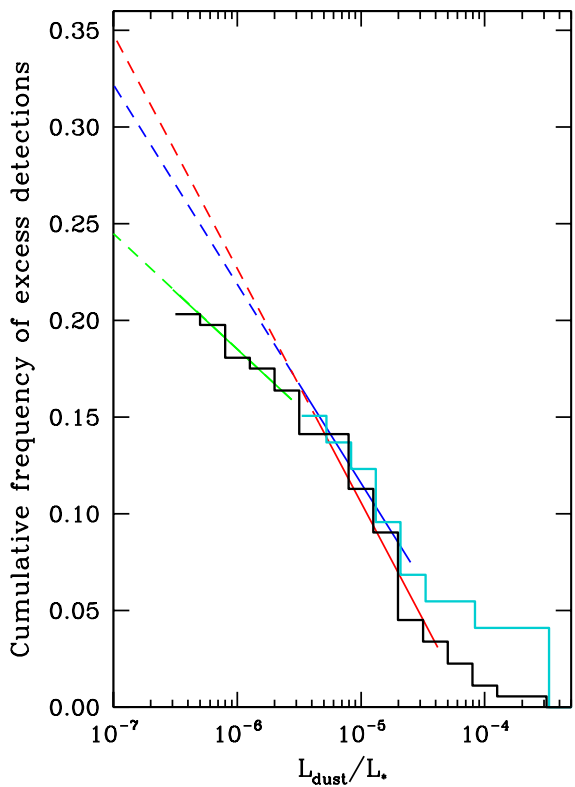}
\caption{Cumulative distributions of the frequency of excesses
  obtained in this work (black) and those obtained from {\it
    Spitzer}/MIPS results \citep{Bryden2006} (cyan), both plotted
  against fractional dust luminosity. The green straight line is a fit
  to the values of the frequencies at the less steep part of our
  cumulative distribution. The extrapolation down to $10^{-7}$, the
  current estimate of the Kuiper-belt fractional luminosity, gives a
  prediction for the incidence rate of excesses of $\sim\!0.25$. The
  extrapolation from a straight line, plotted in red, fitted to the
  steeper part of the distribution, that would mimic the MIPS results,
  gives a higher prediction $\sim\!0.35$. The prediction from
  \cite{Bryden2006} distribution, from the blue straight-line fit,
  is $\sim\!0.32$. See text for details.}
\label{Figure:ldust_cumulative}
\end{figure}

The results on the incidence rates of excesses from PACS data allows
us to push the distribution found from MIPS data, whose minimum
detection limit was $L_{\rm dust}/L_*\!\simeq\!4\times10^{-6}$
\citep{Bryden2006}, down to minimum detections around
$4\times10^{-7}$.  Fig. \ref{Figure:ldust_cumulative} shows a
cumulative distribution, plotted in black, of the frequency of excess
detections from PACS data as a function of the fractional dust
luminosity. The steepness of the distribution decreases at
luminosities lower than $\sim\!4\times10^{-6}$. The extrapolation of a
straight line (in green) fitted to the bins at the left of that value
predicts an incidence of excesses $\sim\!0.25$ at $L_{\rm
  dust}/L_*\!=\!10^{-7}$, the current estimate of the Kuiper-belt dust
fractional luminosity \citep{Vitense2012}.  Plotted in cyan, the
cumulative distribution of the frequency of excesses derived from the
results shown in Tables 2 and 3 of \cite{Bryden2006} (11 excesses out
of 73 stars) is also included in Fig. \ref{Figure:ldust_cumulative};
these were obtained from {\it Spitzer}/MIPS observations at 70 $\mu$m,
with a detection limit of $L_{\rm
  dust}/L_*\!\simeq\!4\times10^{-6}$. The shape of both distributions
is roughly the same in the range
$\sim\!4\times10^{-6}-3\times10^{-5}$. A straight line (in red) fitted
to the bins of our cumulative histogram in that region would predict a
larger incidence of excesses $\sim\!0.35$, at fractional luminosities
of $10^{-7}$. A very similar prediction ($\sim\!0.32$) is obtained
from the cumulative histogram by \cite{Bryden2006}.

The fact that the {\it Spitzer}/MIPS and {\it Herschel}/PACS
distributions are close to each other in the overlapping range of
luminosities is more than a simple consistency check.  The similarity
might indicate that the dust temperature does not correlate with the
dust luminosity in that range.  Indeed, the dust fractional
luminosities in \cite{Bryden2006} were based either on a single 70
$\mu$m measurement or on a combination of 24 $\mu$m and 70 $\mu$m
fluxes.  The dust luminosities in this work were derived from fluxes
at longer wavelengths, basically from 70 $\mu$m, 100 $\mu$m, and 160
$\mu$m PACS fluxes.  For simplicity, let us assume that the Spitzer
luminosities were based solely on a 70 $\mu$m flux and the Herschel
ones on a 100 $\mu$m flux only.  Under this assumption, the fractional
luminosity is directly proportional to the flux at 70 $\mu$m or 100
$\mu$m (see eqn. 3 in \citealt{Bryden2006}).  Next, it is obvious that
the F70/F100 ratio decreases with the decreasing dust temperature.
Should the discs with lower fractional luminosities be systematically
colder, the F70/F100 flux ratio would decrease with the dust
luminosity.  This would make the histogram based on F70 flux flatter
and conversely, the histogram based on the F100 flux steeper.  In the
opposite case, i.e.  if the discs of low fractional luminosity were
systematically warmer, the opposite would be true.  In either case,
the slopes of the two histograms would differ, but this is not what we
see in Fig. \ref{Figure:ldust_cumulative}; instead, they are nearly
the same in the $\sim\!4\times10^{-6}-3\times10^{-5}$ luminosity
range.  Therefore, the dust temperature should be nearly the same for
all discs with the dust luminosities in that range.

What implications might this have? The dust temperature is directly
related to the disc radius.  Assuming that the average stellar
luminosity of all stars with discs in that fractional luminosity range
is nearly the same, the fact that the dust temperature does not depend
on the dust luminosity would automatically mean that the disc radius
does not depend on the dust luminosity either.  This would imply that
it is not the location of the dust-producing planetesimals that
primarily determines the disc dustiness.  Instead, the disc dustiness
must be set by other players in the system, e.g.  by the dynamical
evolution of presumed planets in the disc cavity.

The bottom line of this analysis is that the flattening of the
distribution at low values of $L_{\rm dust}/L_*$, provided by the
results of the {\it Herschel} observations, decreases drastically the
predictions of the excess incidence rate at luminosities $\sim\!10^{-7}$
compared to those obtained from MIPS results. 

\subsection{Debris discs/binarity}

\begin{figure}[!htbp]
\includegraphics[scale=0.65]{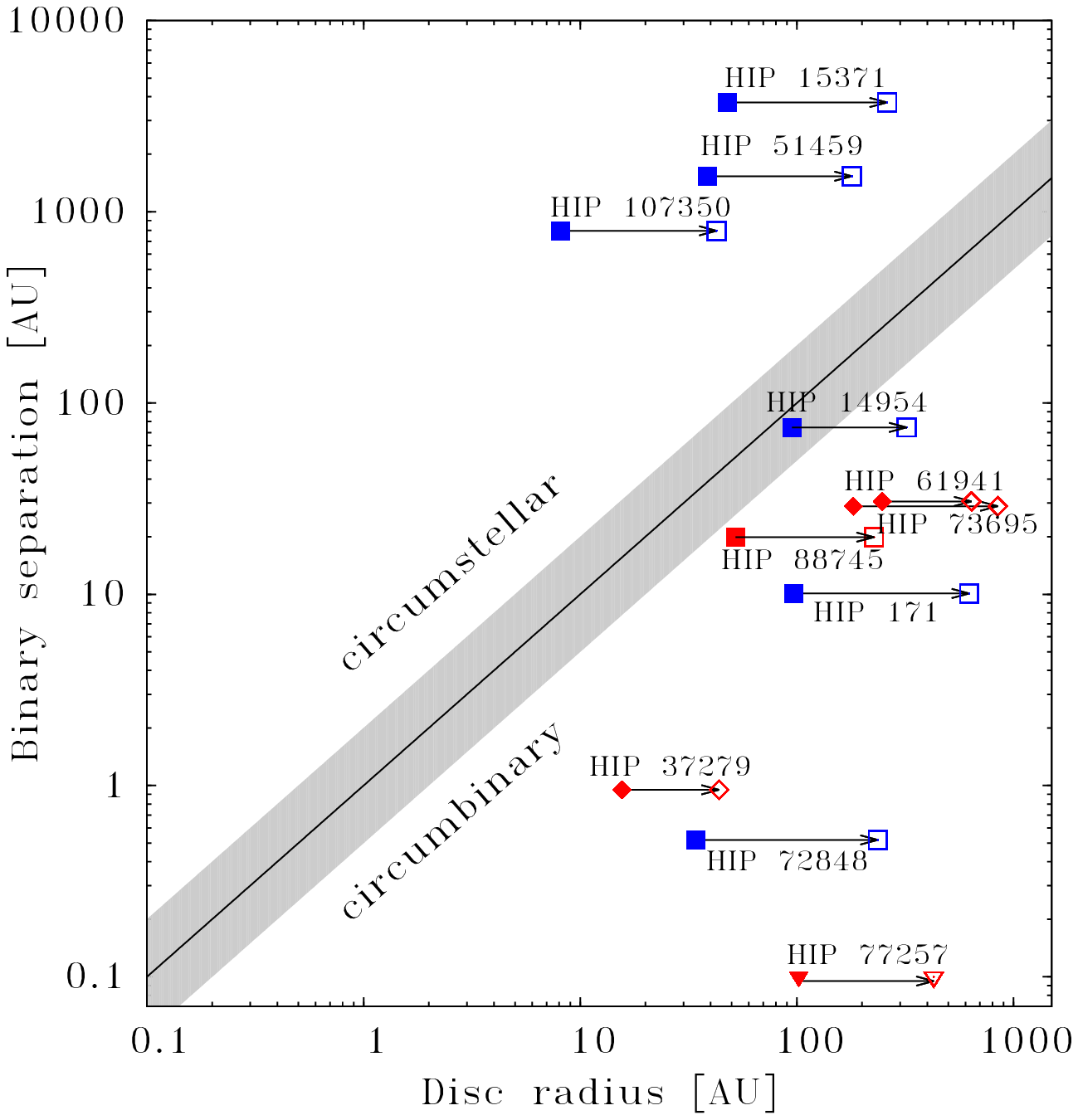}
\caption{Projected binary separation plotted against disc radius for excess 
sources in those stars catalogued as binaries. Red and blue symbols
represent objects studied in this work and in E13, respectively. Squares
and diamonds are confirmed or dubious excesses. Filled and open symbols
represent the value of $R_{\rm BB}$ and $R_{\rm dust}$ (see text for details).
The inverted triangles at the bottom represent the binary HIP 77257 for which 
the separation is not known.}
\label{Figure:binaries}
\end{figure}

Fig. \ref{Figure:binaries} shows the projected binary separation
versus disc radius for excess sources in those stars catalogued as
binaries in Table 4 of this work (red symbols) as well as those in E13
(Table 16 of that paper, blue symbols). Squares and diamonds represent
confirmed and dubious excess sources, respectively; triangles mark the
spectroscopic binary HIP 77257, which has an unknown --but presumably
small-- separation.  Filled and open symbols connected with an arrow
represent the black-body disc radius, $R_{\rm BB}$ computed using
eqn. (\ref{Equation:rbb}) and a more realistic disc radius estimate,
$R_{\rm dust}$, according to \cite{Pawellek2015}, respectively, for
the same system\footnote{Assuming a composition of astrosilicate
  (50\%) and ice (50\%), $\Gamma=5.42\,(L_*/L_\odot)^{-0.35}$, $R_{\rm
    dust} = \Gamma\,R_{\rm BB}$.}.
  
The discs above the diagonal straight line would be circumprimary,
those below the line circumbinary. The grey area roughly marks the
discs that are not expected to exist as they would be disrupted by the
gravity of the companion (assuming the true distance to be close to
the projected one). There is no any apparent reason why all the red
symbols, which correspond to stars in the DUNES\_DB sample, lie in the
``circumbinary'' region. The ranges of the disc radii in DUNES\_DU and
DUNES\_DB are similar, it is only the distributions of the binary
separations that are different (more wide binaries in the DUNES\_DU
sample); whether one finds a wide or a close binary among the excess
sources in one or another sample seems to be purely accidental.

\subsection{Debris discs/metallicity}
\label{Section:metallicity}

\begin{figure}[!htbp]
\includegraphics[scale=0.45]{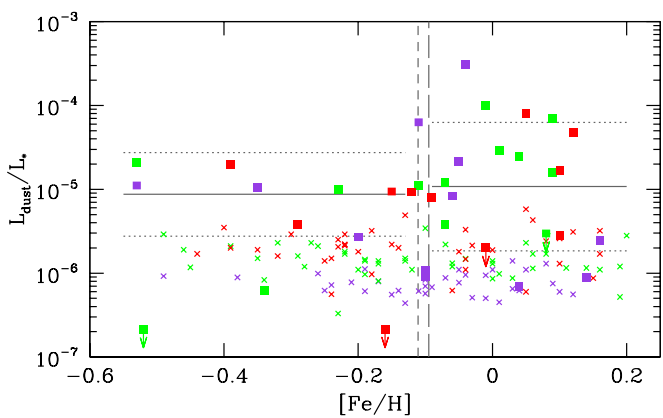}
\caption{Fractional dust luminosities $L_{\rm dust}/L_*$ for the
  DUNES\_DU and DUNES\_DB stars with excesses (squares) and
  $d\!\leq\!20$ pc (Table 14 of E13 and Table \ref{Table:ldust_lstar}
  of this work), and upper limits for the non-excess sources
  (crosses), all of them plotted against metallicity, [Fe/H]. F, G and
  K stars are plotted in violet, green and red. The short-dashed
  vertical line marks the mean metal abundance of the whole sample,
  [Fe/H]=$-0.11$, whereas the long-dashed line marks the median,
  [Fe/H]=$-0.095$. The horizontal lines mark the means (solid) of the
  fractional luminosities for the debris discs of stars showing
  excess, with [Fe/H]$\leq\!-0.11$ and $>\!-0.11$, and plus and minus
  their standard errors (dotted), defined as $\sigma/\sqrt{n}$, where
  $\sigma$ is the standard deviation; these computations have been
  done over the values of $\log (L_{\rm dust}/L_*)$.}
\label{Figure:ldust_metal}
\end{figure}

In Fig. \ref{Figure:ldust_metal}, the fractional luminosities $L_{\rm
  dust}/L_*$ (for the excess sources) and the upper limits (non-excess
sources) for the full DUNES sample with $d\!\leq\!20$ pc, are plotted
against stellar metallicity. It is interesting to note that whereas
above the mean metal abundance of the whole sample, [Fe/H]$_{\rm
  mean}$=$-0.11$, we find a wide range of values of the fractional
luminosity, covering more than two orders of magnitude, the discs
around stars with [Fe/H] below $-0.11$ seem to show a much narrower
interval of $L_{\rm dust}/L_*$, around $\sim\!10^{-5}$. Two
statistical tests performed on the data, namely the
Kolmogorov-Smirnov and the Anderson-Darling, state that the result is
not statistically significant; therefore a larger sample would be
needed to confirm or discard this trend.

\cite{Greaves2006} and \cite{Beichman2006} found that the incidence of
debris discs was uncorrelated with metallicity, these results being
confirmed by \cite{Maldonado2012,Maldonado2015} who found no
significant differences in metallicity, individual abundances or
abundance-condensation temperature trends between stars with debris
discs and stars with neither debris nor planets. However,
\cite{Maldonado2012} pointed out that there could be a deficit of
stars with discs at very low metallicities ($-0.50<{\rm
  [Fe/H]}<-0.20$) with respect to stars without detected discs.
  A recent work by \cite{Gaspar2016}, where the correlation between
  metallicity and debris disc mass is studied, confirmed the finding
  by \cite{Maldonado2012} of a deficit of debris-disc-bearing stars
  over the range $-0.5\!\leq\!{\rm [Fe/H]}\leq\!-0.2$.  Out of the
full sample analysed in this paper, 166 stars have determination of
their metallicities, all of them are plotted in
Fig. \ref{Figure:ldust_metal}; the number of stars with excesses at
both sides of the median metallicity, [Fe/H]$_{\rm median}$=$-0.095$,
is 16 ([Fe/H]$<\!-0.095$) and 20 ([Fe/H]$>\!-0.095$), i.e. an slighly
lower proportion in the low-metallicity side. Considering an average
uncertainty of $\pm0.05$ dex in the metallicities, that could move
objects from one to the other side of the boundary marked by the
median, we can count the number of debris discs around stars with
metallicities lower than [Fe/H]$_{\rm median}\!-\!0.05$ and higher
than [Fe/H]$_{\rm median}\!+\!0.05$, the results being 11/83 and 15/83
objects respectively. The fractions are 0.13$^{+0.09}_{-0.06}$ and
0.18$^{+0.10}_{-0.07}$, respectively, with the uncertainties
corresponding to a 95\% confidence level. Although this result is
  still not statistically significant, it points in the same direction
  as the ones hinted by \cite{Maldonado2012} and confirmed by
  \cite{Gaspar2016}.

\begin{table*}[!htb]
\caption{Statistics of the dust fractional luminosities.}
\label{Table:statistics_ldust}
\begin{tabular}{lcccc}
\hline\hline\noalign{\smallskip}
  &  \multicolumn{3}{c}{$L_{\rm dust}/L_*$} & \\\cline{2-4}\noalign{\smallskip} 
                                   & Mean                   & Median  & MAD & $N$  \\
\hline\noalign{\smallskip}
$\log R'_{\rm HK}>-4.75$ (active)  & $(5.3\pm3.0)\times10^{-5}$ & $1.7\times10^{-5}$  & $1.3\times10^{-5}$    & 10  \\  
$\log R'_{\rm HK}<-4.75$ (inactive)& $(1.8\pm0.6)\times10^{-5}$ & $9.3\times10^{-6}$  & $8.3\times10^{-6}$    & 22  \\[0.25cm]
$P_{\rm rot}<$ 10 d                &  $(5.0\pm3.7)\times10^{-5}$ & $1.2\times10^{-5}$ & $9.7\times10^{-6}$    & 8  \\
10 d $<P_{\rm rot}<$ 30 d          &  $(2.9\pm1.0)\times10^{-5}$ & $1.4\times10^{-5}$ & $1.1\times10^{-5}$    & 12  \\ 
$P_{\rm rot}>$ 30 d                &  $(9.9\pm2.9)\times10^{-6}$ & $8.6\times10^{-6}$ & $4.8\times10^{-6}$    & 8   \\\noalign{\smallskip}
\hline
\end{tabular}
\tablefoot{The uncertainties in the means are the standard errors defined as $\sigma/\sqrt{n}$,
  where $\sigma$ is the standard deviation. MAD (Median Absolute Deviation) is defined as
  MAD={\tt median[abs(x-median(x))]} for a vector {\tt x}.}
\end{table*} 

\subsection{Evolutionary considerations}
\label{Section:evolution}

\begin{figure}[!htbp]
\includegraphics[scale=0.45]{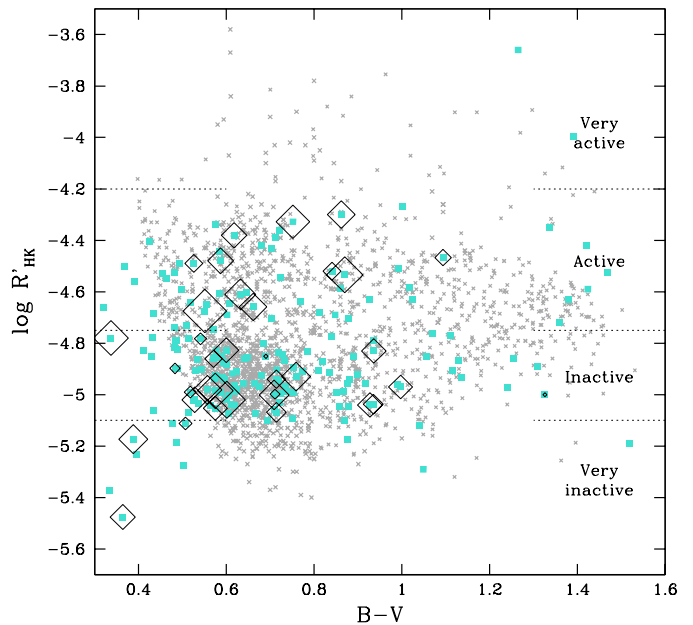}
\caption{A diagram $\log R'_{\rm HK}$ -- $(B\!-\!V)$ showing the stars
  of the full DUNES sample with $d\!\leq\!20$ pc (cyan squares). The
  stars identified to have far-IR excess in E13 and in this paper are
  plotted as diamonds with a size proportional to the value of the
  fractional dust luminosity. In grey, in the background, the samples
  of data by \cite{Henry1996} and \cite{Gray2006}.}
\label{Figure:hkbv}
\end{figure}

Fig. \ref{Figure:hkbv} shows a plot of $\log R'_{\rm HK}$ against
$(B\!-\!V)$ for the DUNES stars within 20 pc, plotted as cyan squares. The
stars with excesses are plotted as diamonds with sizes proportional to
$\log L_{\rm dust}/L_*$.  The regions separating ``very inactive'',
``inactive'', ``active'' and ``very active'' stars, proposed by
\cite{Gray2006} (see their Fig. 4), are also indicated. This diagram
is a qualitative evolutionary picture of FGK main sequence stars in
the sense that as the stars evolve, their rotation rates decrease as a
consequence of angular momentum loss and therefore, their levels of
chromospheric activity also decrease (see e.g. \cite{Skumanich1972},
\cite{Noyes1984}, \cite{Rutten1987}).

Out of the 177 stars of the DUNES sample with $d\!\leq\!20$ pc, 175
have $\log R'_{\rm HK}$ data; 119 objects (25 showing excesses,
21.0\%) are in the region of ``inactive'' or ``very inactive'' stars,
$\log R'_{\rm HK}\!<\!-4.75$, ``old'' objects, whereas 56 (11 with
excesses, 19.6\%) are in the region of the ``active'' or ``very
active'' stars $\log R'_{\rm HK}\!>\!-4.75$, ``young'' objects,
therefore, the incidence of debris discs is similar among inactive or
active stars.

The uneven distribution of stars below and above the activity gap is
more apparent if we use large samples of field FGK stars to check the
relative abundance of active and inactive objects in the solar
neighbourhood. Two catalogues have been used to ascertain this point:
\cite{Henry1996} catalogue contains 814 southern stars within 50 pc;
most of them were chosen to be G dwarfs.  From the \cite{Gray2006}
catalogue, which contains stars within 40 pc, we have retained those
FGK stars with luminosity classes V or V-IV, amounting 1270
objects. The stars of both surveys have been plotted as grey
crosses in Fig. \ref{Figure:hkbv}. The bimodal distribution in stellar
activity first noted by \cite{Vaughan1980} in a sample of northern
objects is seen in both catalogues, the percentage of inactive stars
below the gap frontier, placed at $\log R'_{\rm HK}\!=\!-4.75$ being
73\% \citep{Henry1996}, and 64\% \citep{Gray2006} for
$B\!-\!V\!<\!0.85$. For the sample in this paper, the percentage of
stars below $\log R'_{\rm HK}\!=\!-4.75$ is 68\%.

\begin{figure}[!htbp]
\includegraphics[scale=0.45]{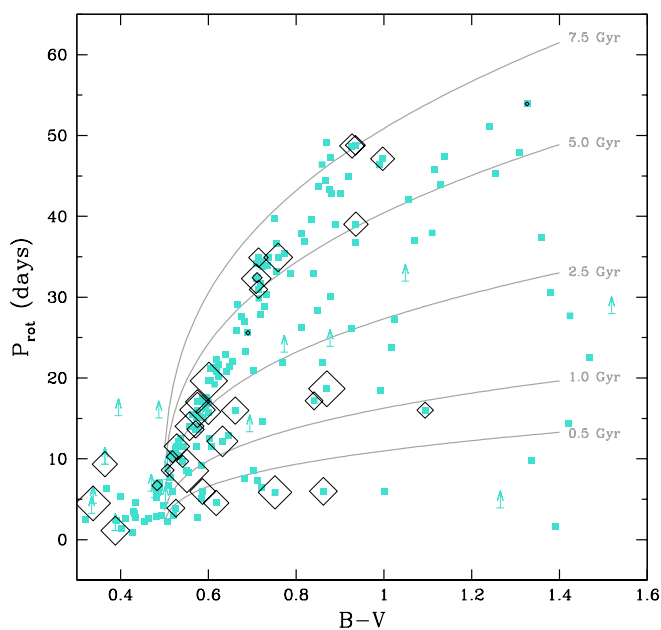}
\caption{A diagram $P_{\rm rot}$ -- $(B\!-\!V)$ showing the stars of
  the full DUNES sample with $d\!\leq\!20$ pc (cyan squares). As in
  Fig. \ref{Figure:hkbv} the stars identified to have far-infrared
  excesses are plotted as diamonds with a size proportional to the
  value of the fractional dust luminosity. Superimposed, five
  ``gyrochrones'' computed following \cite{Mamajek2008} (see Appendix
  A of this work).}
\label{Figure:protbv}
\end{figure}

Fig. \ref{Figure:protbv} shows the rotation periods (see Table 3)
plotted against the $(B\!-\!V)$ colours for the DUNES stars within 20
pc. Five ``gyrochrones'' corresponding to 0.6, 1.0, 2.5, 5.0 and 7.5
Gyr, computed according to \cite{Mamajek2008} are also included in the
graph. The stars with excesses are plotted as black diamonds with
sizes proportional to $\log (L_{\rm dust}/L_*)$. In spite of the
problems involved in the estimation of ages, the reliability of the
computation of rotation periods from the chromospheric activity
indicator $\log R'_{\rm HK}$ (see Appendix A), makes this diagram a
reliable evolutionary scenario of how the sample and the excess
sources are distributed.

In Fig. \ref{Figure:hkbv} it is fairly apparent that the sizes of the
diamonds above the line $\log R'_{\rm HK}\!=\!-4.75$ are, on average,
larger than those below; the line separates active (younger) from
inactive (older) stars. In turn, in Fig. \ref{Figure:protbv} it can be
also seen that the average sizes of the symbols decrease as the
rotation periods increase, therefore suggesting a decreasing
fractional luminosity with increasing stellar age.

To quantify this, we have taken two bins in chromospheric activity,
namely stars above and below $\log R'_{\rm HK}\!=\!-4.75$, and three
bins in rotation period, $P_{\rm rot}\!<\!10$ d, 10 d$<\!P_{\rm
  rot}\!<\!30$ d, and $P_{\rm rot}\!>\!30$ d, and have determined
means and medians of the dust fractional luminosities of the stars in
each bin. The results of this exercise can be seen in Table
\ref{Table:statistics_ldust} where the calculations have been carried
out for the whole sample ($d\!\leq\!20$ pc). Stars with upper limits
in $L_{\rm dust}/L_*$ (three objects in the DUNES\_DU sample and one
in DUNES\_DB) and with lower limits in $P_{\rm rot}$ (four objects)
have not been included; in the last column $N$ specifies the number of
stars used in each case. The effects of decreasing $L_{\rm dust}/L_*$
with decreasing activity and increasing rotation period are
apparent. These results point to the fact that as the stars get older,
there seems to be a slow erosion of the mass reservoir and dust
content in the disc, with the effect of a decrease in $L_{\rm
  dust}/L_*$.

\section{Summary}
\label{Section:summary}

The main goal of this paper --and of the DUNES project-- is to study
the incidence of debris discs around FGK stars in the solar
neighbourhood. Data obtained with the ESA {\it Herschel} space
observatory have been used to characterize the far-IR SED of a sample
of objects observed during two Open Time Key Programmes, namely DUNES
and DEBRIS. A sample of 177 stars within 20 pc has been
analysed. 

\begin{figure}[!htbp]
\includegraphics[scale=0.70]{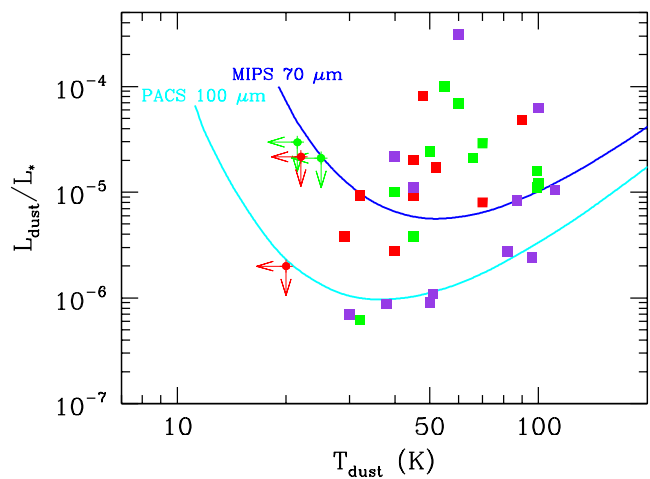}
\caption{A diagram $L_{\rm dust}/L_*$ -- $T_{\rm dust}$ showing the
  position of the 36 stars, out of the 177 in the DUNES sample within
  20 pc, for which an excess has been detected at far-IR wavelengths.
  F, G and K stars are plotted in violet, green and red. The detection 
  limits for a G5 V star at 20 pc for PACS 100 $\mu$m and MIPS 70 $\mu$m
  are also included.}
\label{Figure:ldusttdustsensitivity}
\end{figure}

Fig. \ref{Figure:ldusttdustsensitivity} shows the
fractional dust luminosity, $L_{\rm dust}/L_*$, plotted against the
dust temperature, $T_{\rm dust}$, for the 36 stars for which a far-IR
excess has been detected (see Table 14 of E13 and Table
\ref{Table:ldust_lstar} of this work). The detection limits for a G5 V
star at 20 pc, following \cite{Bryden2006}, are also included in the
graph; the assumed 1$\sigma$ fractional flux accuracies are 20\% for
{\it Spitzer}/MIPS at 70 $\mu$m and 10\% for {\it Herschel}/PACS at
100 $\mu$m (i.e. S/N=10). 14 stars are located below the MIPS 70 $\mu$m
curve, showing how {\it Herschel}/PACS has pushed the limits down to 
fractional luminosities a few times that of the Kuiper-belt.

A summary of the main results follows:

\begin{enumerate}

\item {\it Herschel}/PACS photometry --complemented in some cases with
  SPIRE data-- are provided for the so-called DUNES\_DB sample,
  a set of 54 stars (32 F, 16 G and 6 K) observed by DEBRIS for the
  DUNES consortium. Parameters, ancillary photometry and details on
  the multiplicity are also given. The DUNES\_DU sample was already 
  analysed in detail in E13. See Sect. \ref{Section:sample}, Tables
  2--5, and 7.

\item Eleven sources of the DUNES\_DB sample show excesses in the
  PACS 100 and/or 160 $\mu$m bands (i.e. $\chi_{100}$ and/or
  $\chi_{160}\!>\!3.0$). Five of them are spatially resolved: four
  were previously known, whereas HIP 16852 is a new addition to the
  list of extended sources. This object appears marginally resolved.
  See Tables 7 and \ref{Table:ldust_lstar}, and
  Sect. \ref{Section:excesses}.

\item The DUNES sample --merger of DUNES\_DU and DUNES\_DB-- with
  $d\!\leq\!15$ pc (105 stars) is complete for F stars and almost
  complete for G stars. The number of K stars is large enough to
  provide a solid estimate of the fraction of stars with discs for
  this spectral type. See Sect. \ref{Section:completeness}.

\item The DUNES $d\!\leq\!15$-pc subsample contains 23 F, 33 G and 49
  K stars. The incidence rates of debris discs per spectral type are
  0.26$^{+0.21}_{-0.14}$ (6 objects with excesses out of 23 F stars),
  0.21$^{+0.17}_{-0.11}$ (7 out of 33 G stars) and
  0.20$^{+0.14}_{-0.09}$ (10 out of 49 K stars), the fraction for all
  three spectral types together being 0.22$^{+0.08}_{-0.07}$ (23 out
  of 105 stars). Should $\tau$ Cet and $\epsilon$ Eri --not belonging
  to the DUNES sample-- be included in the statistics, the total
  incidence rate of excesses for stars within $d\!\leq\!15$ pc would
  be $0.23^{+0.09}_{-0.07}$ (25 out of 107 objects). See
  Sect. \ref{Section:incidencerates} and Table
  \ref{Table:incidencerates}.

\item The lowest values of the upper limits for the fractional
  luminosity, $L_{\rm dust}/L_*$, reached are around
  $\sim\!4.0\times10^{-7}$ the median for the whole sample being
  $1.4\times10^{-6}$. These numbers are a gain of one order of
  magnitude compared with those provided by {\it Spitzer} (see
  Sect. \ref{Section:noexcess}). Although it may seem obvious, we must
  stress the fact that the excess detection rates reported in this
  paper are sensitivity limited, therefore, they still represent
  lower limits of the {\it actual} incidence rates. See
  Sect. \ref{Section:incidencerates} and the discussion on
  Fig. \ref{Figure:ldust_cumulative}.

\item There are hints of a different behaviour in the fractional
  luminosities of the discs at lower metallicities and higher
  metallicities, if we split the sample around [Fe/H]$_{\rm
    mean}\!=\!-0.11$: the former seem to cover a narrower interval of
  fractional luminosities than the latter ones.  Splitting around the
  median [Fe/H]]$_{\rm median}\!=\!-0.095$ the sample of stars with
    determinations of metallicities available (166 objects), there
    seems to be a slight deficit of debris discs at lower
    metallicities (16 discs out of 83 stars), when compared to the
    number at higher metallicities (20 discs out of 83 stars). See
    Sect. \ref{Section:metallicity}, and Fig. \ref{Figure:ldust_metal}.

\item Regarding the chromospheric activity, the incidence of debris
  discs is similar among inactive and active stars. There is a decrease
  of the average $L_{\rm dust}/L_*$ with decreasing
  activity/increasing rotation period. Since the stellar activity
  and the spin-down are proxies of the age, this result points to the fact
  that as the stars get older, there seems to be a slow dilution of
  the dust in the disc with the effect of a decrease in its fractional
  luminosity. See Sect. \ref{Section:evolution} and Figs. \ref{Figure:hkbv}
  and \ref{Figure:protbv}.

\end{enumerate}


\begin{acknowledgements}
The authors are gratefuly to the referee for the careful revision of
the original manuscript, comments and suggestions.  We also thank
Francisco Galindo, Mauro L\'opez del Fresno and Pablo Rivi\`ere for
their valuable help.  B. Montesinos and C. Eiroa are supported by
Spanish grant AYA2013-45347-P; they and J.P. Marshall and J. Maldonado
were supported by grant AYA2011-26202. A.V. Krivov acknowledges the
DFG support under contracts KR 2164/13-1 and KR
2164/15-1. J. P. Marshall is supported by a UNSW Vice-Chancellor's
postdoctoral fellowship. R. Liseau thanks the Swedish National Space
Board for its continued support.  A. Bayo acknowledges financial
support from the Proyecto Fondecyt de Iniciación 11140572 and
scientific support from the Millenium Science Initiative, Chilean
Ministry of Economy, Nucleus RC130007. J.-C. Augereau acknowledges
support from PNP and CNES.  F. Kirchschlager thanks the DFG for
finantial support under contract WO 857/15-1. C. del Burgo has been
supported by Mexican CONACyT research grant CB-2012-183007.
\end{acknowledgements}


\begin{appendix}

\section{Rotation periods and ages}

\cite{Noyes1984} studied the Ca {\sc ii}-Rossby number relation for
several sets of mixing-length theory (MLT) models, and found out that
the tightest correlation between $\log R'_{\rm HK}$ and ${\rm
  Ro}\!=\!P_{\rm rot}/\tau_{\rm c}$ was achieved for the set of models
with $\alpha_{\rm MLT}\!=\!1.9$, where $\alpha_{\rm MLT}$ is the ratio
of the mixing length to the pressure scale height in the convection
zone.  $\tau_{\rm c}$, the turnover time, is a characteristic time of
the convection whose precise definition is:

\begin{equation}
\tau_{\rm c}=\frac{H_{\rm p}{\rm (local)}}{v_{\rm local}},
\end{equation}

\noindent where $H_{\rm p}{\rm (local)}$ is the pressure scale height and 
$v_{\rm local}$ is the convective velocity, both quantities computed at a
distance of $0.95\:H_{\rm p}$(base cz) above the base of the outer 
convective zone, where $H_{\rm p}$(base cz) is the pressure scale 
height at the base of the convection zone.

\cite{Noyes1984} derived for that particular value of $\alpha_{\rm MLT}$ 
a polynomial fit for the turnover time as a function of $(B\!-\!V)$. The 
fit is given by the expression:

\begin{equation}
\log \tau_{\rm c}=\left\{
\begin{array}{lr}
1.362-0.166\,x+0.025\,x^2-5.323\,x^3, & x>0  \\
1.362-0.14\,x, & x<0 
\end{array}\right.
\end{equation}

\noindent where $x\!=\!1.0-(B-V)$ and $\tau_{\rm c}$ is given in days.

Once $\tau_{\rm c}$ has been computed, and provided $\log R'_{\rm HK}$
is known, \cite{Noyes1984} gave the following expression:

\begin{equation}
\log {\rm Ro} = 0.324 - 0.400\,y - 0.283\,y^2 - 1.325\,y^3
\end{equation}

\noindent which gives a direct estimation of $P_{\rm rot}$; $y$ is
defined as $y\!=\!5.0\!+\!\log R'_{\rm HK}$.

The determination of $P_{\rm rot}$ from the Rossby number and $\log
R'_{\rm HK}$ is valid if $0.4\!<\!(B\!-\!V)\!<\!1.4$ and $-5.1\!<\!\log
R'_{\rm HK}\!<\!-4.3$ Quantitative similar relationships can be found
in the paper by \cite{Mamajek2008}.

\begin{figure}[!htbp]
\includegraphics[scale=0.45]{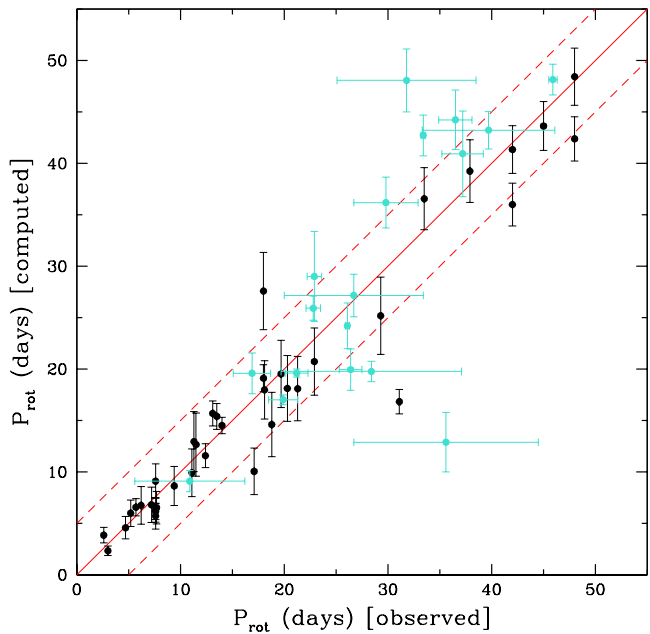}
\caption{Rotation periods estimated according to the formalism
  described in this Appendix, plotted against the rotation periods for
  a sample of FGK stars with known values for $P_{\rm rot}$.  Black
  symbols correspond to stars from \cite{Noyes1984} and cyan symbols
  to stars from \cite{Suarez2015}. The dashed lines mark the interval
  $\pm 5$ days with respect to the diagonal. See text for details.}
\label{Figure:protprot}
\end{figure}

In order to test the validity of these relationships when computing
$P_{\rm rot}$ for the stars analysed in this paper, we have taken the
sample of 39 FGK stars given by \cite{Noyes1984}, and 19 FGK stars by
\cite{Suarez2015}, plus the Sun, for which the rotation period are
known by rotational modulation (first set) and time-series
high-resolution spectroscopy (second set). Fig. \ref{Figure:protprot}
shows the comparison between the observed rotation periods and the
rotation periods estimated following the formalism described above;
the Noyes et al. sample is plotted in black, the Su\'arez-Macare\~no
et al.  sample is plotted in light blue. The uncertainties in both
$(B\!-\!V)$ and $\log R'_{\rm HK}$ have been taken into account when
available. \cite{Noyes1984} do not give any uncertainties in the
observed periods nor in $\log R'_{\rm HK}$, therefore we have
assumed a $\sigma(\log R'_{\rm HK})\!=\!0.05$. It can be seen that the
agreement is fairly good, in particular for the Noyes et
al. sample. Only two objects, namely HD 25171 and HD 40307, are clear
outliers, therefore we can consider the values of $P_{\rm rot}$
computed for the DUNES\_DB stars following this method, and
shown in Table 3 as reliable.

\begin{figure}[!htbp]
\includegraphics[scale=0.45]{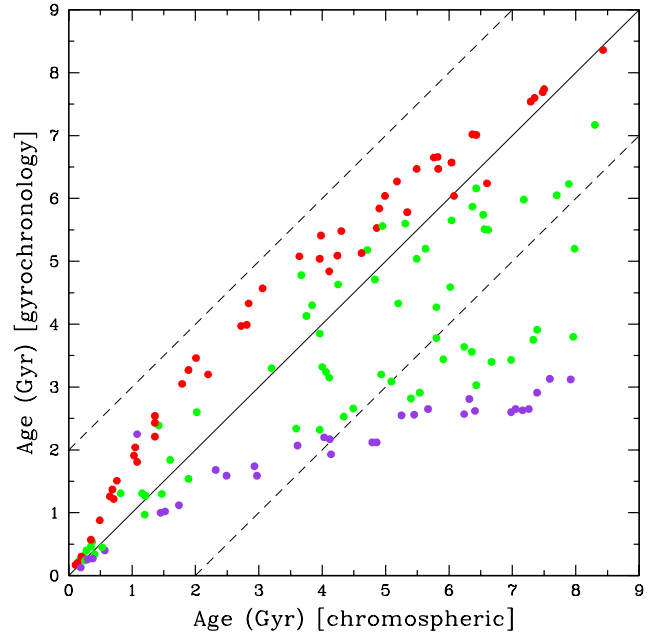}
\caption{Ages for the DUNES\_DU and DUNES\_DB ($d\!\leq\!20$ pc)
  samples estimated by gyrochronology plotted against the
  chromospheric ages. F, G and K stars are plotted in violet,
  green and red, respectively. The dashed lines mark the interval
  $\pm 2$ Gyr with respect to the diagonal.}
\label{Figure:ageage}
\end{figure}

Concerning the ages, all the expressions were extracted from the paper
by \cite{Mamajek2008}. Assuming that $P_{\rm rot}$ is known, equations
(12)--(14) with the revised gyrochronology parameters $a$, $b$, $c$
and $n$ in their Table 10 were used to derive the gyrochronology age
(in Myr):

\begin{equation}
t_{\rm Gyro} = \left(\frac{P_{\rm rot}}{a\,[(B-V)-c]^b}\right)^{1/n}
\end{equation}

\noindent whereas expression (A.3) was used to estimate the chromospheric age:

\begin{equation}
\log t_{\rm HK}= -38.053 - 17.912\,\log R'_{\rm HK}- 1.6675\,(\log R'_{\rm HK})^2
\end{equation}

\noindent where $t_{\rm HK}$ is the age in years. The gyrochronology
expression is valid for $B-V\!>\!0.495$ and the fit to estimate
$t_{\rm HK}$ is only appropriate for $-5.1 < \log R'_{\rm HK}
<-4.0$. Fig. \ref{Figure:ageage} shows a comparison between the ages
of the whole DUNES\_DU plus DUNES\_DB samples estimated using both
approaches; obviously only the stars for which both ages could be
estimated are included in the graph. The discrepancies between the ages 
are clear, the differences being more pronounced for earlier spectral types.

The caveats involving the determination of ages are well known, but
its detailed discussion are out of the scope of this paper (details
can be found in e.g. \cite{Barnes2007}, \cite{Mamajek2008},
\cite{Soderblom2010}, \cite{Epstein2014}, \cite{Meibom2015} and
references therein). The bottom line of this analysis is that whereas
the rotation periods establish a reasonable evolutionary proxy for FGK
stars, in the sense that the older the star the longer the rotation
period, the assignation of an specific age valid in a time span of the
order of Gyr has to be taken with caution.

\section{Spectral energy distributions}

In this appendix the SEDs of the excess, dubious and non excess
sources are shown.

\clearpage

\begin{landscape}
\begin{figure}[!ht]
\includegraphics[scale=0.85]{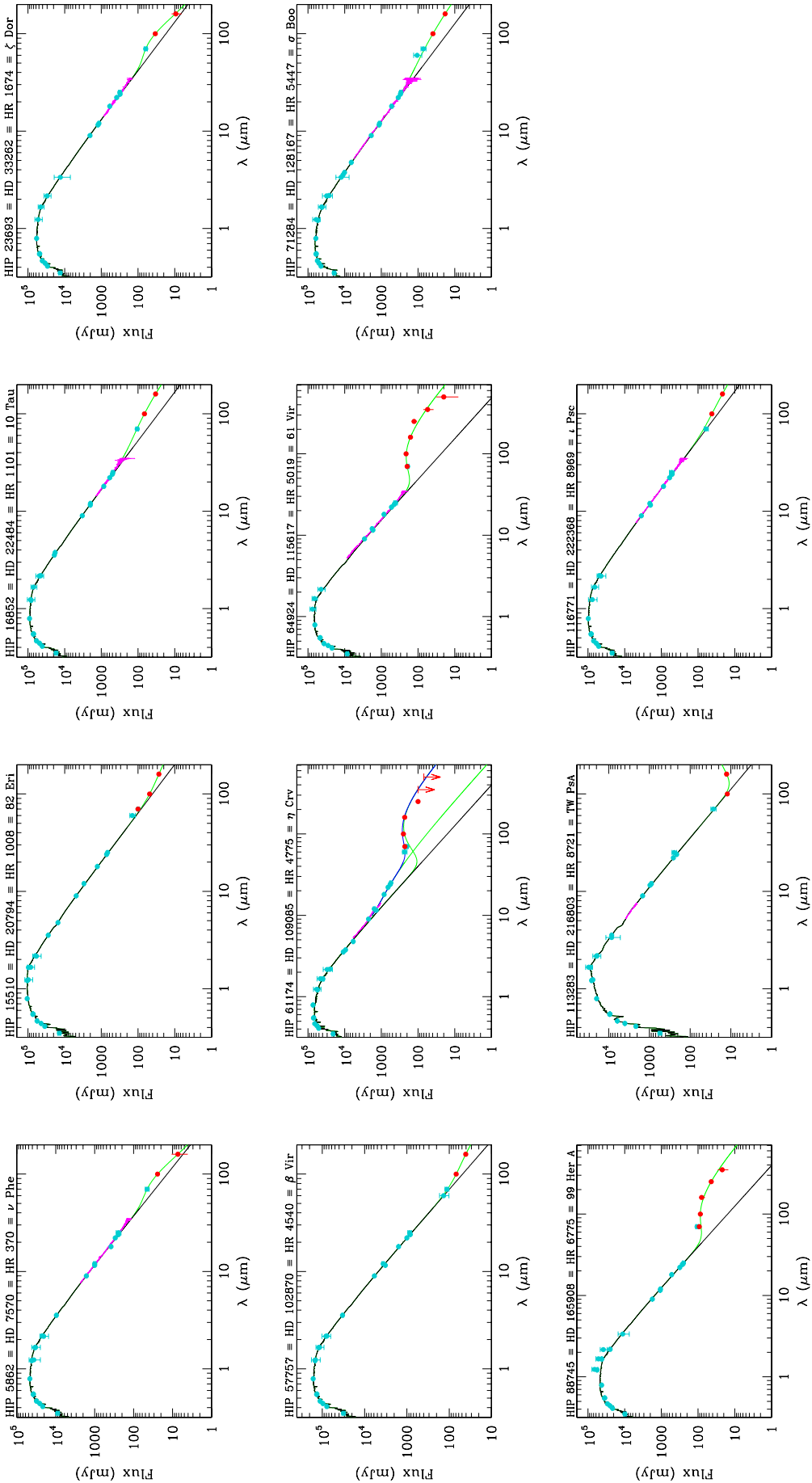}
\caption{SEDs of the FGK DUNES\_DB stars ($d\!\leq\!20$ pc) with excesses.}
\label{Figure:seds_exceses}
\end{figure}
\end{landscape}

\clearpage

\begin{landscape}
\begin{figure}[!ht]
\includegraphics[scale=0.85]{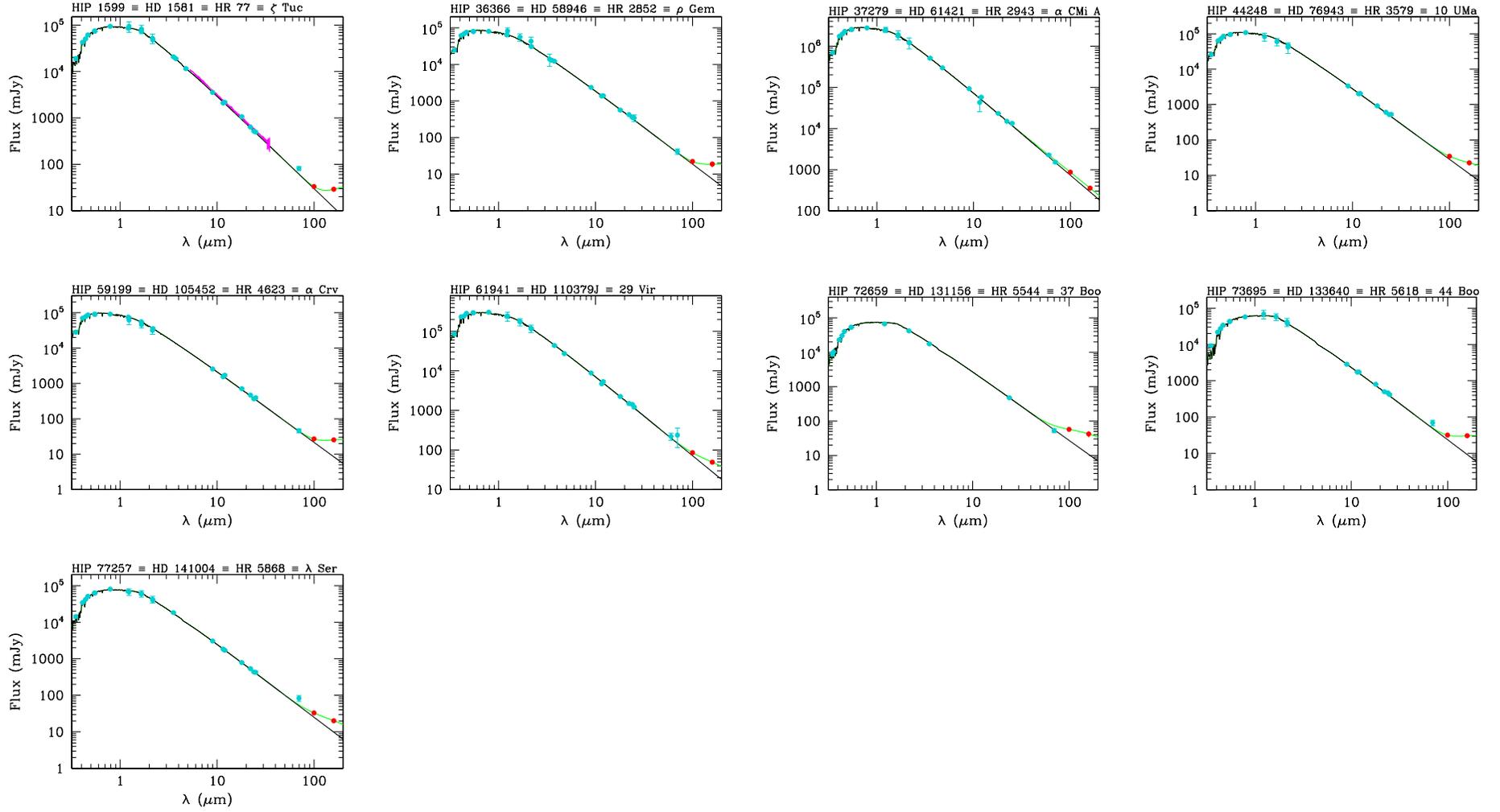}
\caption{SEDs of the FGK DUNES\_DB stars ($d\!\leq\!20$ pc) showing
  apparent excesses but considered dubious.}
\label{Figure:seds_dubious}
\end{figure}
\end{landscape}

\clearpage

\begin{landscape}
\begin{figure}[!ht]
\includegraphics[scale=0.85]{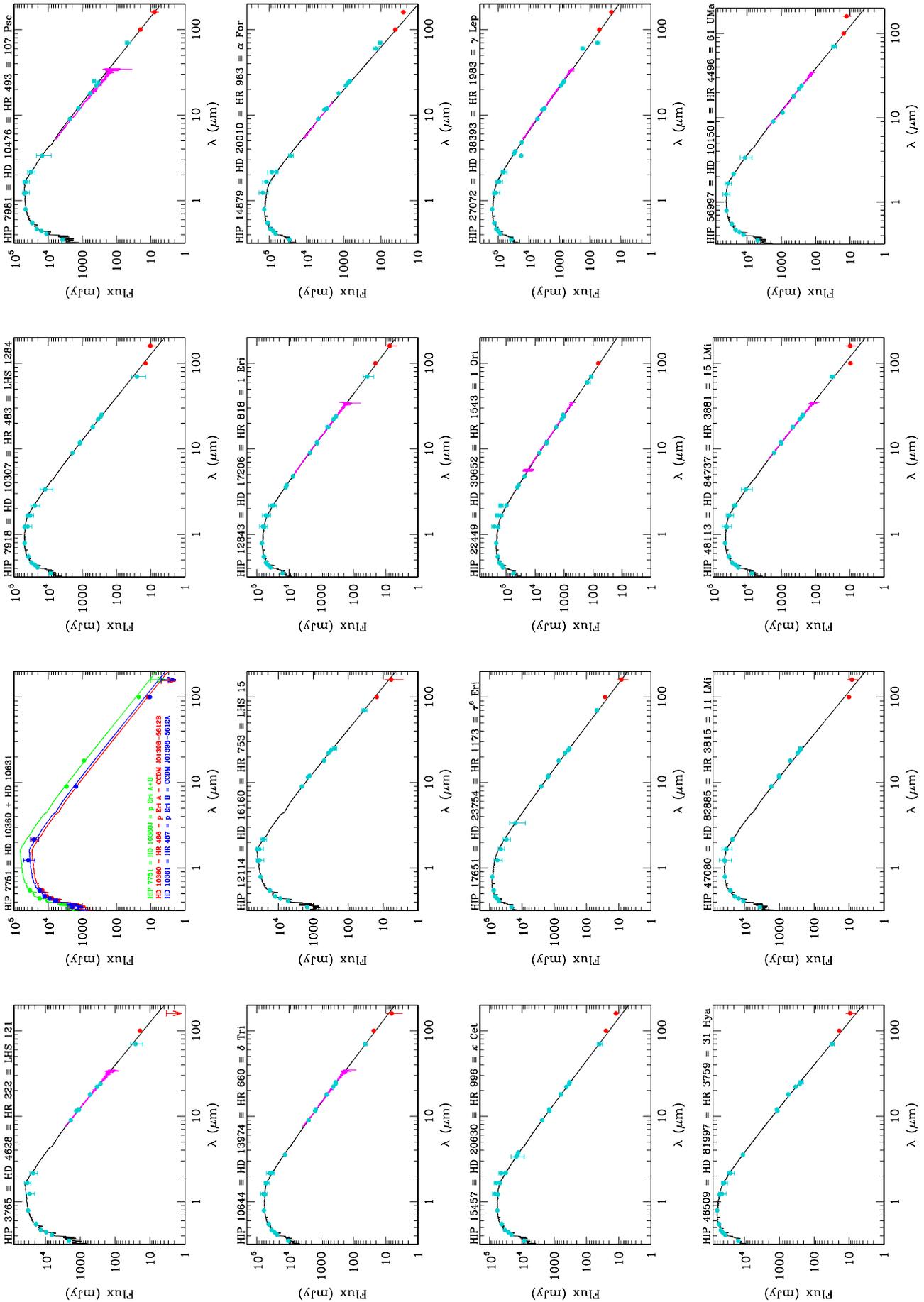}
\caption{SEDs of the FGK DUNES\_DB stars ($d\!\leq\!20$ pc) without excess.}
\label{Figure:seds_noexcess}
\end{figure}
\end{landscape}

\clearpage

\addtocounter{figure}{-1}
\begin{landscape}
\begin{figure}[!ht]
\includegraphics[scale=0.85]{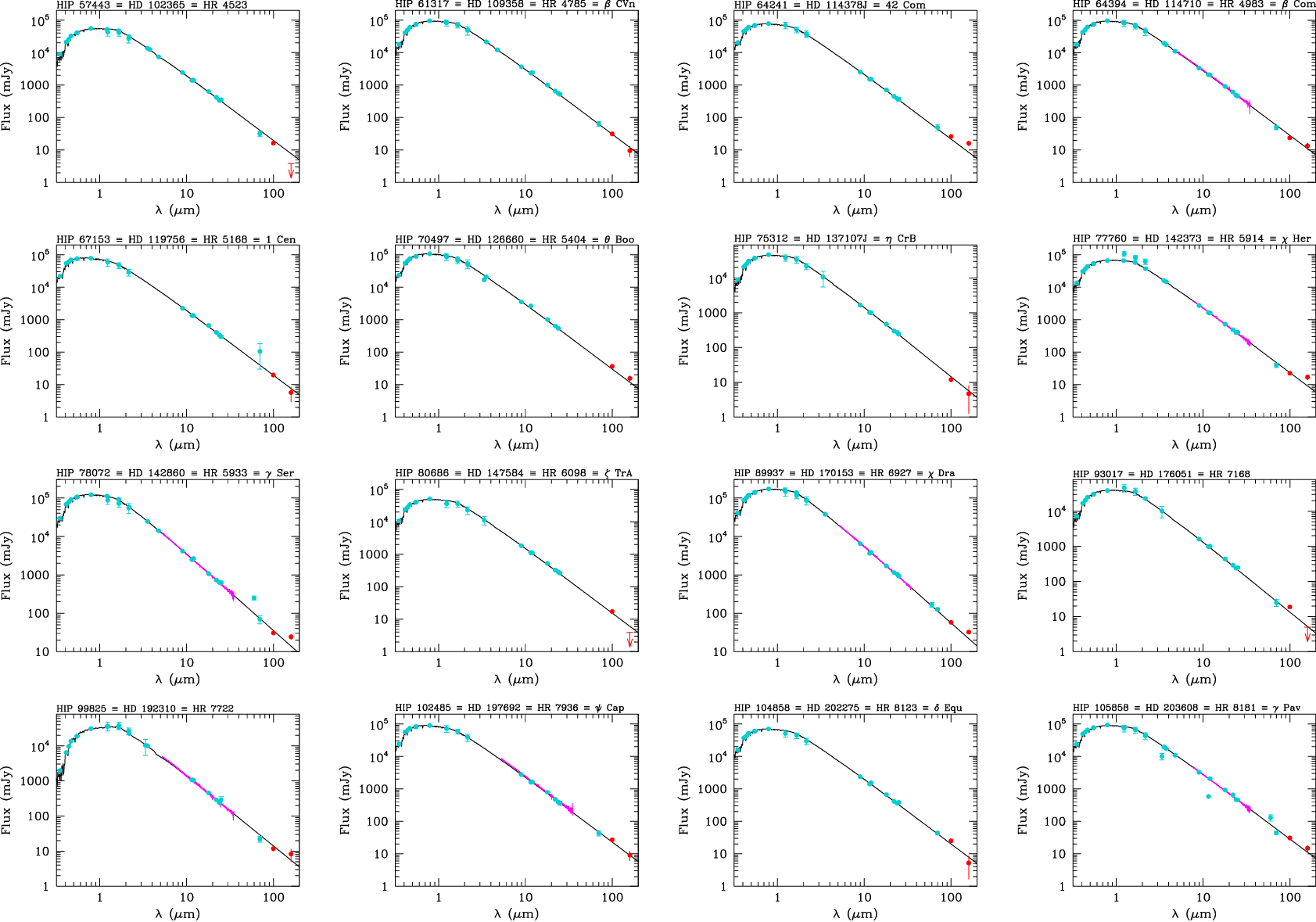}
\caption{Continued.}
\end{figure}
\end{landscape}

\clearpage

\addtocounter{figure}{-1}
\begin{landscape}
\begin{figure}[!htbp]
\includegraphics[scale=0.85]{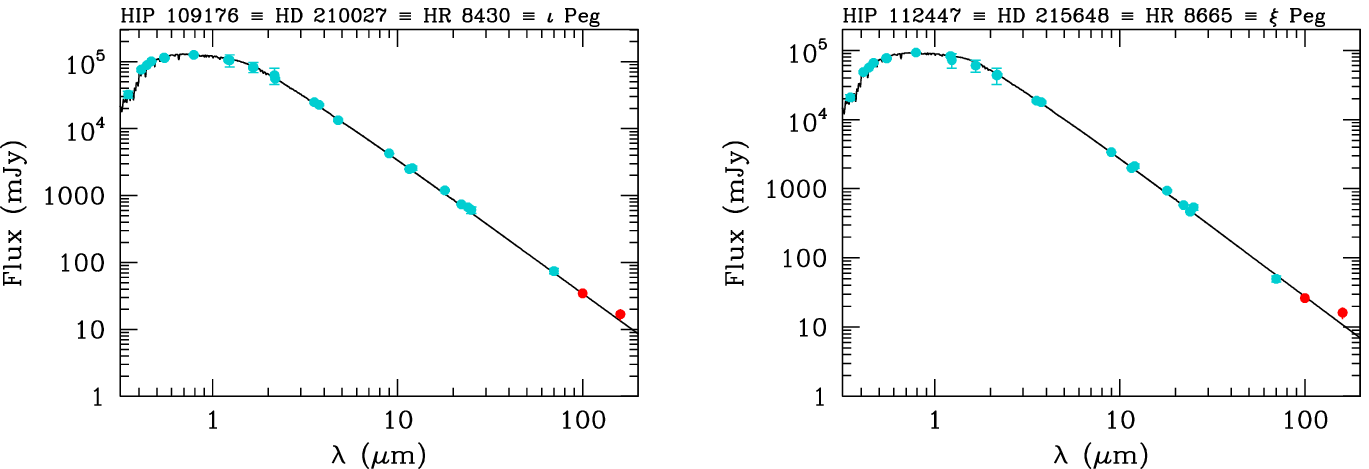}
\caption{Continued.}
\end{figure}
\end{landscape}

\end{appendix}

%
\setcounter{table}{1}

\onecolumn
\begin{scriptsize}
\begin{longtable}{lllllllrr}
\multicolumn{9}{c}{{\bf \tablename\ \thetable{}.} The subsamples of DUNES\_DB with $d\!\leq\!15$ pc and $15\!\,\,{\rm pc}\!<\!d\!\leq\!20$ pc.}\\
\\
\hline\hline
HIP & HD & Other ID & SpT (Hipparcos) & SpT range & \multicolumn{2}{c}{ICRS (2000)} & \multicolumn{1}{c}{$\pi$(mas)} & \multicolumn{1}{c}{$d$(pc)} \\
\hline
\endfirsthead
\multicolumn{9}{c}{{\bf \tablename\ \thetable{}.} Continued.}\\
\\
\hline\hline
HIP & HD & Other ID & SpT (Hipparcos) & SpT range & \multicolumn{2}{c}{ICRS (2000)} & \multicolumn{1}{c}{$\pi$(mas)} & \multicolumn{1}{c}{$d$(pc)}  \\
\hline
\endhead
\hline
\multicolumn{9}{l}{Continued\dots}\\
\endfoot
\endlastfoot 
\multicolumn{9}{l}{ }\\                                                                                                                          
\multicolumn{9}{l}{FGK stars ($d\!\leq\!15$  pc)} \\\hline
1599    &  1581    &  $\zeta$ Tuc       & F9V      & F9.5V, F9V--G0V             & 00 20 04.260 & $-$64 52 29.25 &  116.46$\pm$0.16 &    8.59$\pm$0.01 \\       
3765    &  4628    &  LHS 121           & K2V      & K2.5V, K2V                  & 00 48 22.977 & $+$05 16 50.21 &  134.14$\pm$0.51 &    7.45$\pm$0.03 \\       
7751    &  10360J  &  p Eri             & K0V      & K0V, K1--K3V                & 01 39 47.540 & $-$56 11 47.10 &  127.84$\pm$2.19 &    7.82$\pm$0.13 \\    
7918    &  10307   &  HR 483            & G2V      & G1.5V, G1V                  & 01 41 47.143 & $+$42 36 48.12 &   78.50$\pm$0.54 &   12.74$\pm$0.09 \\        
7981    &  10476   &  107 Psc           & K1V      & K1V, K0V--K1V               & 01 42 29.762 & $+$20 16 06.60 &  132.76$\pm$0.50 &    7.53$\pm$0.03 \\     
10644   &  13974   &  $\delta$ Tri      & G0V      & G0.5V, F9V--G0V             & 02 17 03.230 & $+$34 13 27.23 &   92.73$\pm$0.39 &   10.78$\pm$0.05 \\
\end{longtable}
\end{scriptsize}


\newpage
%
\setcounter{table}{3}

\begin{table*}
\scriptsize
\begin{center}
\begin{tabular}{lcccccccccc}
\multicolumn{11}{c}{{\bf \tablename\ \thetable{}.} Parameters of the DUNES\_DB stars with $d\!\leq\!15$ pc and $15\!\,\,{\rm pc}\!<\!d\!\leq\!20$ pc.}\\
\\
\hline\hline
HIP & $T_{\rm eff}$ & $\log g$ & [Fe/H] & Flag &  $L_{\rm bol}$ & $\log R'_{\rm HK}$ & Ref & $P_{\rm rot}$ (d)  & $t_{\rm Gyro)}$ & $t_{\rm HK}$\\
    &  (K)        & (cm/s$^2$) & (dex) &      & ($L_\odot$)   &                   &     & or lower limit\dag& (Gyr)            &  (Gyr)        \\
\hline
\hline
\\
\multicolumn{11}{l}{ }\\
\multicolumn{11}{l}{FGK stars ($d\!\leq\!15$  pc)} \\\hline                 
1599    & 5960 & 4.45 & $-0.17$ & S  & 1.224  & $-4.855$ & 1 & $14.01\pm0.97$        & 2.20   & 4.03   \\
\end{tabular}
\end{center}
\end{table*}

\newpage
%
\setcounter{table}{4}
\begin{table*}
\scriptsize
\begin{center}
\begin{tabular}{lllllrrl}
\multicolumn{8}{c}{{\bf \tablename\ \thetable{}.} Multiplicity and planetary companions of the DUNES\_DB stars with $d<20$ pc.}\\
\\
\hline
\hline
HIP &   Multiplicity   & CCDM identification & \multicolumn{2}{c}{ICRS (2000)}    &  $\mu_\alpha\cos(\delta)$ & $\mu_\delta$ & Separation \\
    &   and details    &   & \multicolumn{1}{c}{$\alpha$ (hh mm ss)} & \multicolumn{1}{c}{$\delta$ (deg min sec)} & (mas/yr) & (mas/yr) & (arcsec)   \\
\hline
\multicolumn{8}{l}{ }\\    
\multicolumn{8}{l}{FGK stars ($d\!\leq\!15$  pc)} \\\hline

7751   & Double or multiple star          &                    &              &                 &&&                         \\
       & p Eri B K2V  V=5.797             & CCDM J01398-5612A  & 01 39 47.557 &  $-$56 11 47.34 &  286.10 & 16.16 & \multirow{2}{*}{11.18}  \\
\end{tabular}
\begin{tabular}{lllclc}
\hline\hline\noalign{\smallskip}
HIP    & Comment                   & CCDM identification & \multicolumn{3}{c}{Planets} \\\cline{4-6}\noalign{\smallskip}
       &                           &                     &   &  M/M$ _{\rm Jup}$   &  a (UA)  \\\hline\noalign{\smallskip}
15510  & High proper-motion star   &                     & b & 0.009   & 0.12 \\
       &                           &                     & c & 0.008   & 0.20 \\
       &                           &                     & d & 0.015   & 0.35 \\\noalign{\smallskip}
57443  & High proper-motion star   & CCDM J11465-4030A   & b & 0.05    & 0.46 \\\noalign{\smallskip}
64924  & High proper-motion star   & CCDM J13185-1818A   & b & 0.016   & 0.05 \\
       &                           &                     & c & 0.057   & 0.22 \\
       &                           &                     & d & 0.072   & 0.48 \\\noalign{\smallskip}
93017  & Spectroscopic binary      &                     & b & 1.5     & 1.76 \\\noalign{\smallskip}
99825  & Variable star             &                     & b & 0.053   & 0.32 \\
       &                           &                     & c & 0.076   & 1.18 \\
\noalign{\smallskip}\hline\noalign{\smallskip}
\multicolumn{6}{l}{All planets discovered by the radial velocity (RV) method except HIP 93017b (astrometry),}\\
\multicolumn{6}{l}{therefore masses for the RV exoplanets are lower limits.}\\
\multicolumn{6}{l}{Data from http://exoplanetarchive.ipac.caltech.edu/docs/data.html except for HIP 93017b,}\\
\multicolumn{6}{l}{which are taken from http://exoplanet.eu}
\label{ }
\end{tabular}
\end{center}
\end{table*}

%
\setcounter{table}{5}
\begin{table*}
\scriptsize
\begin{center}
\begin{tabular}{lrlllll}
\multicolumn{7}{c}{{\bf \tablename\ \thetable{a}.} Johnson $V$, $B\!-\!V$, Cousins $V\!-\!I$ and Str\"omgren photometry.}\\
\\
\hline
\hline
HIP  &  \multicolumn{1}{c}{$V$} & \multicolumn{1}{c}{$B\!-\!V$} & \multicolumn{1}{c}{$V\!-\!I$} & \multicolumn{1}{c}{$b\!-\!y$} &  \multicolumn{1}{c}{$m_1$} & \multicolumn{1}{c}{$c_1$} \\   
     &  \multicolumn{1}{c}{(mag)} & \multicolumn{1}{c}{(mag)} & \multicolumn{1}{c}{(mag)} & \multicolumn{1}{c}{(mag)} &  \multicolumn{1}{c}{(mag)} & \multicolumn{1}{c}{(mag)} \\
\hline
       &      &                 &               &                 &                 &                  \\
\multicolumn{7}{l}{FGK stars ($d\!\leq\!15$ pc)} \\\hline
  1599 & 4.23 & 0.576$\pm$0.010 & 0.65$\pm$0.02 & 0.368$\pm$0.003 & 0.177$\pm$0.004 & 0.302$\pm$0.020  \\
  3765 & 5.74 & 0.890$\pm$0.008 & 0.97$\pm$0.02 & 0.512$\pm$0.003 & 0.423$\pm$0.002 & 0.255$\pm$0.004  \\
\end{tabular}
\label{ }
\end{center}
\end{table*}

\setcounter{table}{5}
\begin{table*}
\scriptsize
\begin{center}
\begin{tabular}{lrrrcrrrrrrl}
\multicolumn{12}{c}{{\bf \tablename\ \thetable{b}.}  2MASS $JHK_{\rm s}$ and ancillary Johnson $JHKLL'M$ photometry.}\\
\\
\hline
\hline
HIP  &  \multicolumn{1}{c}{2MASS $J$} & \multicolumn{1}{c}{2MASS $H$} & \multicolumn{1}{c}{2MASS $K_{\rm s}$} &  Qflag & \multicolumn{1}{c}{$J$} & \multicolumn{1}{c}{$H$} & \multicolumn{1}{c}{$K$} &  \multicolumn{1}{c}{$L$} & \multicolumn{1}{c}{$L'$} & \multicolumn{1}{c}{$M$} & Refs.      \\   
     &  \multicolumn{1}{c}{(mag)} & \multicolumn{1}{c}{(mag)} & \multicolumn{1}{c}{(mag)} &     & \multicolumn{1}{c}{(mag)} & \multicolumn{1}{c}{(mag)} & \multicolumn{1}{c}{(mag)} &  \multicolumn{1}{c}{(mag)} & \multicolumn{1}{c}{(mag)} & \multicolumn{1}{c}{(mag)} &    \\
\hline
       &                   &                  &                  &          &        &        &        &        &        &        &     \\
\multicolumn{12}{l}{FGK stars ($d\!\leq\!15$ pc)} \\\hline
  1599 &   3.068$\pm$0.272 &  2.738$\pm$0.218 &  2.769$\pm$0.250 & DDD      &  3.196 &  2.880 &  2.832 &  2.803 &  2.795 &  2.856 & 1,2,3 \\
  3765 &   4.367$\pm$0.310 &  3.722$\pm$0.230 &  3.683$\pm$0.268 & DDD      &        &        &        &        &        &        &     \\
\end{tabular}
\end{center}
\end{table*}

\setcounter{table}{5}
\begin{table*}
\scriptsize
\begin{center}
\begin{tabular}{lcccccc}
\multicolumn{7}{c}{{\bf \tablename\ \thetable{c}.}  AKARI 9 and 18 $\mu$m fluxes and WISE W1, W3 and W4 photometry.}\\
\\
\hline
\hline
HIP   & \multicolumn{2}{c}{AKARI}   & &     \multicolumn{3}{c}{WISE}         \\\cline{2-3}\cline{5-7} 
      & 9  $\mu$m & 18  $\mu$m      & & 3.35 $\mu$m (W1) & 11.56 $\mu$m (W3) & 22.09 $\mu$m (W4) \\
      & (mJy)      & (mJy)          & & (mag)            &  (mag)            &  (mag)            \\
\hline
       &                          &                         &   &                   &                  &                      \\
\multicolumn{7}{l}{FGK stars ($d\!\leq\!15$ pc)} \\\hline
1599   &   4193$\pm$28     &  1055$\pm$30    &   &                   &  2.856$\pm$0.011 &   2.788$\pm$0.019    \\
\end{tabular}
\end{center}
\end{table*}

\setcounter{table}{5}
\begin{table*}
\scriptsize
\begin{center}
\begin{tabular}{ll@{ }rl@{ }rl@{ }rc@{}cc}
\multicolumn{10}{c}{{\bf \tablename\ \thetable{d}.} IRAS 12, 25, 60 $\mu$m and {\it Spitzer}/MIPS 24 and 70 $\mu$m fluxes.}\\
\\
\hline
\hline
HIP   &\multicolumn{6}{c}{IRAS} & & \multicolumn{2}{c}{MIPS}  \\\cline{2-7}\cline{9-10}
      & 12 $\mu$m& \%   & 25 $\mu$m & \% & 60 $\mu$m & \%    & & \multicolumn{1}{c}{24 $\mu$m}  & \multicolumn{1}{c}{70 $\mu$m} \\
      & \multicolumn{1}{c}{(mJy)} & &  \multicolumn{1}{c}{(mJy)} & & \multicolumn{1}{c}{(mJy)}& & & \multicolumn{1}{c}{(mJy)}  & \multicolumn{1}{c}{(mJy)}  \\
\hline
       &          &     &          &      &          &     & &                       &                       \\
\multicolumn{7}{l}{FGK stars ($d\!\leq\!15$ pc)} \\\hline
1599   & 3.11$\times10^3$ &  4  & 6.95$\times10^2$ &   4  &                 &     & & 519$\pm$11   & 82   $\pm$8   \\
\end{tabular}
\end{center}
\end{table*}


%
\setcounter{table}{6}
\begin{table*}
\scriptsize
\begin{center}
\begin{tabular}{llcccc}
\multicolumn{6}{c}{{\bf \tablename\ \thetable{}.} OBSIDs for the DUNES\_DB stars with $d\!\leq\!20$ pc.}\\
\\
\hline\hline
HIP & HD & PACS & Scan & X-Scan & On-source time [s] \\
\hline
\multicolumn{6}{c}{ }\\
\multicolumn{6}{l}{FGK stars ($d\!<\!15$ pc)}\\
\hline
1599                  & 1581 & 100/160 & 1342198529 & 1342198530 & 288 \\ 
3765                  & 4628 & 100/160 & 1342213215 & 1342213216 & 288 \\\noalign{\vspace{0.05cm}}
\multirow{2}{*}{7751} & 10360 & 100/160 & 1342193169 & 1342193170 & 288 \\
\hline
\end{tabular}
\end{center}
\end{table*}

%
\setcounter{table}{7}
\begin{table*}
\scriptsize
\begin{center}
\begin{tabular}{llrrrrrrl}
\multicolumn{9}{c}{{\bf \tablename\ \thetable{}.} PACS flux densities for the DUNES\_DB stars with $d\!\leq\!15$ pc and $15\!\,\,{\rm pc}\!<\!d\!\leq\!20$ pc.}\\
\\
\hline
\hline
HIP & HD & \multicolumn{1}{c}{PACS100} &  \multicolumn{1}{c}{S100}  & $\chi_{100}$ & \multicolumn{1}{c}{PACS160} & \multicolumn{1}{c}{S160} & $\chi_{160}$ & Status \\ 
    &    &  \multicolumn{1}{c}{(mJy)}  &  \multicolumn{1}{c}{(mJy)} &              & \multicolumn{1}{c}{(mJy)}   & \multicolumn{1}{c}{(mJy)} &             &        \\
\hline
\multicolumn{9}{l}{ }\\                                                         \multicolumn{9}{l}{FGK stars ($d\!\leq\!15$  pc)} \\\hline
1599                  &  1581    &  33.23$\pm$ 2.35 & 30.08$\pm$ 0.19  & $ 1.34$ &   29.03$\pm$3.46  &  11.75$\pm$ 0.07 & $ 4.99$ & Dubious    \\
3765                  &  4628    &  19.43$\pm$ 1.93 & 15.84$\pm$ 0.36  & $ 1.83$ &          $<3.38$  &   6.19$\pm$ 0.14 & $     $ & No excess \\
\end{tabular}
\end{center}
\end{table*}

\end{document}